\documentclass[11pt,a4paper]{article}
\pdfoutput=1

\usepackage{xcolor}
\usepackage{jheppub}
\usepackage{amsfonts}
\usepackage{amsmath}
\usepackage{pifont}
\usepackage{slashed}
 
\newcommand{\eq}[2]{\begin{equation} #1 \label{#2} \end{equation}}

\newcommand{\Tr}{{\rm Tr}\, \,}

 \DeclareMathOperator{\extdm}{d}
 \newcommand{\extd}{\extdm \!}
\newcommand{\gaij}[1]{\ga_{ij}^{(#1)}}
\newcommand{\beij}[1]{\be_{ij}^{(#1)}}
\newcommand{\erho}[1]{e^{#1 \rho}}

\newcommand{\gij}[1]{g_{ij}^{(#1)}}
\newcommand{\bij}[1]{b_{ij}^{(#1)}}
\newcommand{\hij}{h_{ij}^{(0)}}

\newcommand{\bup}[1]{b^{ij}_{(#1)}}
\newcommand{\hup}{h^{ij}_{(0)}}

\newcommand{\al}{\alpha}
\newcommand{\be}{\beta}
\newcommand{\ga}{\gamma}
\newcommand{\de}{\delta}
\newcommand{\si}{\sigma}
\newcommand{\la}{\lambda}

\newcommand{\eps}{\epsilon}

\newcommand{\ka}{\kappa}

\newcommand{\Ga}{\Gamma}
\newcommand{\La}{\Lambda}
\newcommand{\Om}{\Omega}

\newcommand{\cG}{{\cal G}}
\newcommand{\cO}{{\cal O}}

\subheader{} %% just for drafts

\title{Holographic two-point functions for 4d log-gravity}

\author[a]{Niklas Johansson}
\author[b,c]{\!\!,~Ali Naseh}
\author[a,d]{and Thomas Zojer}
\affiliation[a]{Institute for Theoretical Physics, 
           Vienna University of Technology,\\
           Wiedner Hauptstr. 8--10/136,
           A-1040 Vienna, Austria.}
\affiliation[b]{Department of Physics, Sharif University of Technology,\\
                P.O.~Box 11365-9161, Tehran, Iran\\
                and \\
                School of Physics, Institute of Research in Fundamental Sciences (IPM),\\
                P.O.~Box 19395-5531, Tehran, Iran.}
\affiliation[c]{Institute for Theoretical Physics, University of Amsterdam,\\
     Science Park 904, Postbus 94485, 1090 GL Amsterdam, The Netherlands.}
\affiliation[d]{Centre for Theoretical Physics, University of Groningen,\\
                Nijenborgh 4, 9747 AG Groningen, The Netherlands.}

\emailAdd{niklasj@hep.itp.tuwien.ac.at, naseh@ipm.ir, t.zojer@rug.nl}

\abstract{We compute holographic one- and two-point functions of critical higher-curvature gravity in four dimensions. 
The two most important operators are the stress tensor and its logarithmic partner, sourced by ordinary 
massless and by logarithmic non-normalisable gravitons, respectively. In addition, the logarithmic gravitons
source two ordinary operators, one with spin-one and one with spin-zero. The one-point function of the 
stress tensor vanishes for all Einstein solutions, but has a non-zero contribution from logarithmic gravitons.
The two-point functions of all operators match the expectations from a three-dimensional logarithmic conformal field theory.
}

\arxivnumber{1205.5804}


\begin{document}

\begin{flushright} TUW-12-09 \\ UG-12-12 \\ IPM/P-2012/021 \end{flushright}

\maketitle

\flushbottom 

\section{Introduction}

During the last couple of years, a lot of effort has been put into the understanding of higher-curvature gravity theories in
three space-time dimensions. In these studies, one centre of attention has been so-called ``critical tunings'' in the space
of coupling constants. The corresponding theories exhibit several features making them potentially interesting as candidates 
for models of quantum gravity.

The first instance of this criticality phenomenon was unravelled in cosmological topologically massive gravity (TMG) \cite{Deser:1981wh,Deser:1982vy}.
The action of TMG consists of an Einstein--Hilbert term, a cosmological constant and a gravitational Chern--Simons term. In
a seminal paper, Li, Song and Strominger \cite{Li:2008dq} noted that there is a certain tuning of the Chern--Simons coupling
$\mu$ where several interesting 
phenomena simultaneously occur. First, one of the central charges of the boundary CFT vanishes and second, 
the linearised equations of motion degenerate.

The fact that the central charge is zero means that, if the theory is unitary, it has to be chiral. 
Therefore, the partition function would be trivially holomorphic, fulfilling the hope \cite{Witten:2007kt} apparently 
not realised \cite{Maloney:2007ud} for pure AdS gravity. This was taken as a strong
indication that TMG at the critical point may have a consistent quantum mechanical formulation.

In the same vein, the degeneration of the linearised equations of motion was taken as evidence that the massive graviton
--- a negative energy state for general tunings --- is absent in the critical case. However, soon after the chiral gravity
conjecture, it was realised that even at the critical point there is a physical bulk mode $\psi^{\rm log}$ with negative
bulk energy \cite{Grumiller:2008qz}. As in logarithmic CFT \cite{Gurarie:1993xq}, the Hamiltonian was shown to have a Jordan
block structure on the Hilbert space including $\psi^{\rm log}$ and based upon this it was conjectured \cite{Grumiller:2008qz}
that TMG at the critical point is holographically dual to a logarithmic CFT.
 
The logarithmic mode has a different asymptotic behaviour than other modes, it diverges more quickly. This is
not a problem in itself --- the variational principle is well-defined, and the required boundary conditions 
are consistent \cite{Grumiller:2008es}. However, it opens up for the possibility to
eliminate the mode by imposing certain boundary conditions\footnote{We note in passing that a similar idea was advocated
not long ago to eliminate the intrinsic ghosts of conformal gravity in four dimensions \cite{Maldacena:2011mk}.}. It was
subsequently shown \cite{Maloney:2009ck}, that the asymptotic behaviour cannot be avoided by constructing wave-packets with
compact support because of a linearisation instability.

There are therefore dual reasons to focus some attention on critical
points in the study of higher-curvature gravity. On the one hand, they offer a potential mechanism eliminating problematic
negative energy states. On the other hand --- keeping the logarithmic modes --- they offer holographic descriptions of
logarithmic CFTs: an AdS/LCFT correspondence.

As its discovery, the first technical pieces of evidence for this correspondence were provided 
in the context of TMG. Two-point functions were computed according to (suitably generalised) standard AdS/CFT methods \cite{Skenderis:2009nt}, resulting in perfect agreement with LCFT correlators, and these results were generalised to three-point functions in a technical {\it tour de force} \cite{Grumiller:2009mw}. The one-loop partition function was
also demonstrated to match LCFT expectations \cite{Gaberdiel:2010xv}. These developments have since been generalised to various 
other versions of higher-curvature gravity in three dimensions exhibiting critical points. 
(For an extensive list of references see, e.g., \cite{Grumiller:2010rm, Grumiller:2010tj,  Bertin:2011jk}.)

In particular, a certain combination of higher-curvature terms results in a theory that propagates
unitary gravitons \cite{Bergshoeff:2009hq, Bergshoeff:2009aq}. 
This theory, which has acquired the increasingly misleading name ``new massive gravity'' (NMG),
also allows for critical tunings leading to logarithmic behaviour \cite{Grumiller:2009sn, Alishahiha:2010bw,
Gaberdiel:2010xv}. 

Naturally, it is of interest to lift this entire discussion to higher dimensions.
Such a venture was initiated by Lu and Pope \cite{Lu:2011zk} who formulated a higher-curvature gravity theory,
being a four-dimensional analogue of critical NMG. This model was later generalised to arbitrary dimensions \cite{Deser:2011xc}. 
Subsequent works \cite{Alishahiha:2011yb,Gullu:2011sj,Bergshoeff:2011ri} have demonstrated and categorised the logarithmic excitations in the theory.
However, the logarithmic structure is so far explored to a far lesser extent than in the aforementioned lower-dimensional literature. 
For instance, there are no results on any linearisation instability (potentially allowing for elimination of the 
logarithmic modes), there are no results on partition functions, and no correlators have been computed. 

In this work we take a first step toward a more complete higher-dimensional AdS/LCFT correspondence: we compute the
one- and two-point functions of several operators in four-dimensional critical
gravity. These operators are the stress tensor, its logarithmic partner and two other operators corresponding to the
transverse vector and scalar pieces of the logarithmic graviton. 
All results are completely consistent with an LCFT$_3$ dual. Our computations follow to a large extent
the lead of \cite{Skenderis:2009nt} which also contains clear and comprehensible explanations of the 
associated logic and technicalities. 

This paper is organised as follows. In section \ref{sec:first} we briefly review critical gravity in four dimensions
and derive the first variation of the action. In section \ref{sec:second} we obtain the renormalised second variation of the 
action. This requires the construction and evaluation of several holographic counterterms.
In section \ref{sec:correlators} we explicitly calculate the linearised modes and use them to derive 
the two-point correlators for critical gravity, and in section \ref{seq:concl} we conclude. 
Three appendices are devoted to technicalities that are not
presented in the main text.

\section{First variation of the action}\label{sec:first}

To compute one- and two-point functions according to the AdS/CFT dictionary \cite{Maldacena:1997re, Gubser:1998bc, Witten:1998qj}, one functionally differentiates the on-shell action with respect to the relevant sources. In this section we perform the first variation of the
action and compute the corresponding boundary term. This object will be differentiated further in later sections.
We also evaluate the first variation with respect to the boundary metric allowing to determine the one-point function
of the stress tensor. The computation of one-point functions in general requires holographic renormalisation, and the
method we follow was developed in \cite{Henningson:1998gx, Henningson:1998ey}. (For reviews, see 
\cite{Skenderis:2002wp, Papadimitriou:2004ap}.)

Our model is the one presented in \cite{Lu:2011zk} with bulk action
\eq{
I_{\rm bulk} = \frac{1}{2\ka^2} \int \extd^4 x \, \sqrt{-g} \,  \left(R - 2\La + \al R^{\mu\nu}R_{\mu\nu} + \be R^2 \right) \, .
}{eq:action}
Such higher-curvature actions generally propagate a massless spin-two graviton, a massive spin-two field and a massive scalar \cite{Stelle:1976gc, Stelle:1977ry}. The two latter excitations are ghosts. If the parameters $\al$ and $\be$ are tuned as $\al = -3\be$, the spin-zero excitation is absent. 
If furthermore $\be = -1/2\La$ the theory becomes critical: the black holes have zero mass and entropy 
and the massive graviton becomes logarithmic \cite{Lu:2011zk, Alishahiha:2011yb, Bergshoeff:2011ri}.

Varying \eqref{eq:action} with respect to the metric $g_{\mu\nu}$ produces
\eq{
\de I_{\rm bulk} = \frac{1}{2\ka^2} \int \extd^4 x \, \sqrt{-g} \, \big( {\rm EOM}^{\mu\nu}\de g_{\mu\nu} 
+ \nabla_\si J^\si \big) \,,
}{eq:var1}
where
\begin{align}\label{eq:var2}
&{\rm EOM}^{\mu\nu} = -\cG^{\mu\nu} - E^{\mu\nu}  \,,\\ \label{eq:var3}
&J^{\si} =  A^{\mu\nu} \de \Ga^{\si}_{\mu\nu} - A^{\mu\si} \de \Ga^{\la}_{\la\mu} 
+ \left( \frac{1}{2} \nabla_\la A^{\si\la} g^{\mu\nu} - \nabla^\mu A^{\si \nu} + \frac{1}{2}\nabla^\si A^{\mu\nu} \right) \de g_{\mu\nu}\, ,
\end{align}
where the tensor $A^{\mu\nu}$ is defined as
\eq{
A^{\mu\nu} = (1+2\be R)g^{\mu\nu} + 2\al R^{\mu\nu}\, .
}{}
The tensors appearing in the equations of motion are furthermore
\begin{align} \label{eq:G}
\cG_{\mu\nu}= & \,R_{\mu\nu} -\frac12 R g_{\mu\nu} + \La g_{\mu\nu} \,, \\ 
\begin{split}
E_{\mu\nu} =&  \,2\al (R_{\mu\la}R_{\nu}^\la  - \frac14 R^{\la\si} R_{\la\si}g_{\mu\nu} ) + 2\be R (R_{\mu\nu} - \frac14 R g_{\mu\nu})\\
&+ \al(\Box R_{\mu\nu} + \frac12\Box R g_{\mu\nu} - 2\nabla_\la \nabla_{(\mu}R_{\nu)}^\la ) + 2\be(g_{\mu\nu}\Box R - 
\nabla_{\mu}\nabla_{\nu} R) \, .
\end{split} \label{eq:E}
\end{align}
We shall from here on only consider the case $\al = -3\be$. To reduce clutter we also fix $\La = -3$.
Using the (twice contracted) Bianchi identity it is easy to show that, for $\al = -3\be$, the tensor
$E_{\mu\nu}$ is traceless. Thus, taking the trace of the equations of motion then establishes
\eq{
R = 4\La = -12\,.
}{}
In particular the Ricci scalar is constant on-shell, eliminating several of the terms in Eq.~\eqref{eq:E}.
Using again the Bianchi identities the equations of motion simplify to
\eq{
\cG_{\mu\nu} + \frac{3\be}{2} R^{\la\si} R_{\la\si} g_{\mu\nu} - 24\be (R_{\mu\nu} + 3 g_{\mu\nu})
 -3\be (\Box R_{\mu\nu} + 2 R_{\al\mu\la\nu} R^{\al \la}) = 0\, .
}{}
Note that these are the correct equations of motion only in vacuum. Coupling the theory to matter would
of course require keeping the full result \eqref{eq:G}--\eqref{eq:E}.

Let us now fix Gaussian normal coordinates,
\eq{
\extd s^2 = \extd \rho^2  + \ga_{ij}(\rho)\extd x^i \extd x^j \, .
}{eq:FG1}
From \eqref{eq:var1} it is then clear that the on-shell variation reads
\eq{
\de I_{\rm bulk} |_{\rm EOM} = \frac{1}{2\ka^2}\int_{\partial M} \extd^3 x\, \sqrt{-\ga} \, J^{\rho}  \,,
}{eq:onshell}
with $J^{\mu}$ defined in Eq.~\eqref{eq:var3}. 
Partially integrating the Christoffel symbols in $J^{\rho}$ allows us to write the variation as
\eq{
\begin{split}
\de I_{\rm bulk} |_{\rm EOM} = \frac{1}{2\ka^2}&\int_{\partial M} \extd^3 x\, \sqrt{-\ga}  \, J^{\rho}
 = \int_{\partial M} \extd^3 x\, \sqrt{-\ga}\, \Big( -\big(A^{ij} + A^{\rho\rho}\ga^{ij}\big)\de K_{ij} \\
&+ \left[\frac{1}{2}\nabla_{\rho}(A^{ij} + A^{\rho\rho}\ga^{ij}) + \nabla_k A^{\rho k}\ga^{ij} - \nabla^i A^{\rho j} + A^{\rho\rho}K^{ij} \right]
\de \ga_{ij}
\Big)\, .
\end{split}
}{eq:1stvarMain}
This is the expression that we shall differentiate further to obtain the two-point correlators in later sections. In doing so, 
we shall use Poincar\'{e} patch AdS$_4$ as a background, to obtain the correlators corresponding to a CFT$_3$ on a flat background.
This computation will also give us the one-point functions on a flat background.

Let us first, however, compute the one-point function of the stress tensor in global AdS. The advantage of this is that we can
then obtain the conserved Poincar\'{e} charges of, e.g., a black hole.
Our background metric is thus global AdS$_4$:
\eq{
\extd s^2 = -\cosh^2\! \rho \extd t^2 + \extd \rho^2 + \sinh^2\! \rho \extd \Om_2^2  \,,
}{}
and the matrix $\ga_{ij}$ in Eq.~\eqref{eq:FG1} is assumed to have the expansion
\eq{
\begin{split}
\ga_{ij} = \ga^{(0)}_{ij} e^{2\rho} + \ga^{(2)}_{ij} -  \be^{(3)}_{ij} \, \rho \, e^{-\rho} + \ga^{(3)}_{ij} e^{-\rho} + \ldots 
%= \\
%         &= \ga^{(0)}_{ij} \frac{1}{y^2} + \ga^{(2)}_{ij} +  \be^{(3)}_{ij} \,y \, \log y  + \ga^{(3)}_{ij} y + \ldots
\end{split}
}{eq:FG2}
with the leading contributions fixed
\eq{
\ga^{(0)}_{ij} = \begin{pmatrix} -1/4 & 0 & 0 \\ 0 & 1/4 & 0  \\ 0 & 0 & \frac{1}{4}\sin^2 \theta  \end{pmatrix}\, , \qquad
\ga^{(2)}_{ij} = \begin{pmatrix} -1/2 & 0 & 0 \\ 0 & -1/2 & 0  \\ 0 & 0 & -\frac{1}{2}\sin^2 \theta  \end{pmatrix}\, .
}{eq:FG3}
The term $\ga^{(3)}_{ij}$ corresponds to the massless graviton and the term $\be^{(3)}_{ij}$ is forced to vanish by the equations 
of motion except at the critical point $\be = 1/6$, where it captures the logarithmic mode. For other tunings, the massive graviton has 
a different power law fall-off. We ignore such terms for brevity, but still keep $\beta$ arbitrary since it illuminates some of the
results. It is important to keep in mind though, that our result for the stress tensor below is incomplete unless $\be = 1/6$.

Using these expansions it is straightforward to obtain an expansion for the variation in Eq.~\eqref{eq:onshell}.
This computation is detailed in appendix \ref{FG} and the result is
\eq{
\de I_{\rm bulk} |_{\rm EOM} = \frac{1}{2\ka^2}\int_{\partial M} \extd^3 x\, \sqrt{-\ga}
   \left([1-6\be](K^{ij}\de \ga_{ij} - 2 \ga^{ij}\de K_{ij} )  - \frac{27\be}{2} \erho{-5} \be^{ij}_{(3)} \de \ga_{ij}\right)\, . 
   }{eq:deI}
Here $K_{ij}$ is the extrinsic curvature
\eq{
K_{ij} = \frac{1}{2}\partial_{\rho} \ga_{ij} \, .
}{}
The first term in this variation is divergent and contains variations of $\ga^{(3)}_{ij}$ and $\be^{(3)}_{ij}$, destroying a well-defined variational principle. These terms must in general be cancelled by holographic counterterms. Note however, that at the critical point $\be = 1/6$ 
the first term vanishes. Thus, there is no need to add holographic counterterms for the critical case.
This is in complete analogy with what happens in three dimensions 
for new massive gravity \cite{Hohm:2010jc,Alishahiha:2010bw}.

Off the critical locus, counterterms are needed. It is shown in appendix \ref{FG} that the required
terms are exactly the ones for pure Einstein gravity \cite{Balasubramanian:1999re,deHaro:2000xn}\footnote{For a less directly
comparable, but earlier, computation of the counterterms for Einstein gravity, see
\cite{Henningson:1998gx, Henningson:1998ey}.}, multiplied with the appropriate prefactor:
\eq{
I_{\partial M} = - \frac{1-6\be}{2\ka^2} \int_{\partial M} \extd^3 x \, \sqrt{-\ga} \, \Big(4-2K+R[\ga] \Big) \, .
}{eq:Ict}
The boundary stress tensor $T^{ij}$ is defined as
\eq{
\de I_{\rm ren} |_{\rm EOM} = \frac{1}{2} \int_{\partial M} \extd^3 x\, \sqrt{-\ga^{(0)}}\,\,T^{ij}\,\de\ga_{ij}^{(0)} \, ,
}{eq:ST}
or, shorter,
\eq{
T_{ij} = -\frac{2}{\sqrt{-\ga^{(0)}}}\frac{\de I_{\rm ren} |_{\rm EOM}}{\de\ga^{ij}_{(0)}} \,,
}{eq:ST2}
where $I_{\rm ren} = I_{\rm bulk} + I_{\partial M}$.
At the critical point the stress tensor takes the simple form
\eq{
T_{ij}^{{\rm crit}}=-\frac{9}{4 \ka^2}\,\be_{ij}^{(3)} \, ,
}{eq:tijcrit}
while for generic values of $\be$ the result is
\eq{
T_{ij}=\big(1-6\be \big)\,\frac{3}{2 \ka^2}\, \ga_{ij}^{(3)} \, .
}{eq:tij}
The latter result of course contains contributions from the boundary action in Eq.~\eqref{eq:Ict}, and, for $\beta=0$,
is consistent with \cite{deHaro:2000xn}. We note again
that the result \eqref{eq:tij} only contains the contribution from solutions captured by $\ga^{(3)}_{ij}$, e.g.,
massless gravitons and black holes.

As shown in appendix \ref{FG}, the asymptotic equations of motion imply
\eq{
\Tr \be^{(3)} \equiv \ga^{ij}_{(0)}\be^{(3)}_{ij} = \Tr \ga^{(3)} \equiv \ga^{ij}_{(0)}\ga^{(3)}_{ij} = 0 \,,
}{eq:traceless}
and 
\eq{
\nabla_{(0)}^k \be_{ki}^{(3)} = 0\, ,   \qquad \qquad \Big(1-6\be \Big) \, \nabla_{(0)}^k \ga_{ki}^{(3)}=0\, ,
}{eq:divergence}
where the covariant derivative $\nabla_{(0)}$ is taken with respect to the boundary metric $\ga^{(0)}_{ij}$.
These equations imply that the boundary stress tensor --- for general $\be \neq 1/6$ at least the part we computed --- 
is traceless and conserved. Note that $\gamma^{(3)}_{ij}$ is not necessarily transverse for the critical case.
The non-transverse components make up part of the logarithmic graviton.

Our result \eqref{eq:tijcrit} shows that any solution having a vanishing $\be^{(3)}_{ij}$ also has a vanishing stress tensor.
In particular this means that any solution to Einstein gravity has vanishing mass and angular momentum, confirming the
result of \cite{Lu:2011zk} for the mass. The situation is completely parallel to that in NMG \cite{Clement:2009gq}. 

\section{Second variation of the action} \label{sec:second}

To be able to compute the two-point correlators we first need the second variation of the action, or, equivalently,
the first variation of the one-point functions. This entails computing the first variation of the coefficients
multiplying $\delta \gamma_{ij}$ and $\delta K_{ij}$
in \eqref{eq:1stvarMain}. To technically simplify this computation we shall perform it perturbatively around
Poincar\'e patch AdS as opposed to the global metric. This means that the boundary is a plane, and that we
obtain planar CFT correlation functions.

Thus, we consider a metric of the form
\eq{
\extd s^2 = \frac{\extd y^2}{y^2} + \ga_{ji}\extd x^i \extd x^j =
 \frac{\extd y^2 + \eta_{ij}\extd x^i \extd x^j}{y^2} + h_{ij}\extd x^i \extd x^j  \,,
}{}
where $\eta_{ij}$ is the flat Minkowski metric on the boundary $\mathbb{R}^{2,1}$. 
The perturbation $h_{ij}$ is assumed to have a Fefferman--Graham expansion of the form
\eq{
h_{ij} = b^{(0)}_{ij}\frac{\log y}{y^2} + h^{(0)}_{ij} \frac{1}{y^2} + b^{(2)}_{ij}\log y +  g^{(2)}_{ij}
+ b^{(3)}_{ij}\, y \, \log y +  g^{(3)}_{ij}\, y + \ldots 
}{eq:FGh}
To compute the correlator of the logarithmic mode we must include the leading logarithmic term $b^{(0)}_{ij}$ that breaks
the asymptotic AdS property of the metric. By the equations of motion, this also requires the inclusion of the term
$b^{(2)}_{ij}\log y$. Furthermore, the equations of motion require $\bij{0}$ be traceless.

In AdS/CFT language, the expansion coefficients $h^{(0)}_{ij}$ and $b^{(0)}_{ij}$ represent sources for different operators.
The former is a source for the stress-energy tensor ($T_{ij}$), whereas the latter contains sources for its logarithmic
partner, but also
for other operators. Collectively, we shall denote these operators by $t_{ij}$. The tracelessness of $\bij{0}$ carries
over to the operator(s) $t_{ij}$ and imposes the constraint that $t_{ij}$ be traceless too.
Correlators of these operators are
then given by functionally differentiating the on-shell action with respect to $h^{(0)}_{ij}$ and $b^{(0)}_{ij}$.

Our normalisation of these operators will be defined by
\begin{subequations}
\begin{align}
 \langle T_{ij} \rangle &= 
 2 \frac{\de I_{\rm ren} |_{\rm EOM}}{\de h^{ij}_{(0)}}\,, \qquad \qquad \langle T_{ij} \, ... \rangle 
 = -2 i \frac{\de }{\de h^{ij}_{(0)}}  \langle ... \rangle \label{eq:T} \,,\\
 \langle t_{ij} \rangle &=   2 \frac{\de I_{\rm ren} |_{\rm EOM}}{\de b^{ij}_{(0)}}\,, \qquad \qquad\, \langle t_{ij} \, ... \rangle 
 = -2 i \frac{\de }{\de b^{ij}_{(0)}}  \langle ... \rangle \label{eq:t} \,,
  \end{align}
\end{subequations}
where the ellipsis denotes any operator. The factor of $-i$ in the two rightmost expressions comes about because the generating function is actually $\sim i I_{\rm ren}$ (for Lorentzian signature),
as explained in appendix B of \cite{Skenderis:2009nt}. 

\subsection{Second variation of the bulk action}

The computation of the second variation of the bulk action is lengthy but straightforward. The details and the conventions 
are presented in appendix \ref{app:2nd}. The final result is
\eq{
\begin{split}
\de^{(2)}I_{\rm bulk}|_{\rm EOM} &= \frac{1}{2\kappa^2}\int\limits_{\partial M} \extd^3 x \sqrt{-\gamma}\, 
      \Big\{ -\frac{3}{4} \de \bup{0} \de \bij{0} \frac{1}{y^3} 
+ \Big[2\de\hup \de\bij{2} -\frac{1}{2}\de \bup{0}\de \bij{2}- \\
&- 2 \de\bup{0}\de\gij{2} \Big] \frac{1}{y} -\frac{9}{2} \de\bup{0}\de\bij{3} \log y - \frac{9}{4}\de\bup{0}\de\gij{3} - \frac{9}{4}\de\hup \de\bij{3}\, \Big\} \,,
\end{split}
}{eq:2ndvar}
where all indices are raised by $\eta^{ij}$. We also have already symmetrised\footnote{
This means, e.g., setting $\de_1 \bup{0}\de_2 \bij{2} + \de_2 \bup{0}\de_1 \bij{2} = 
2 \de \bup{0}\de \bij{2}$ and setting $\de_1\bup{0}\de_2\bij{3} -  \de_2\bup{0}\de_1\bij{3}  = 0$ and so on.} 
in the two variations, since only the symmetrised version is needed in computing the correlators. 

We see that setting $b^{(0)}_{ij}$ and $\bij{2}$ to zero recovers our old result \eqref{eq:tijcrit} for the 
stress tensor at the critical point. Furthermore, because of the many divergent terms it is clear that the computation
of two-point correlators requires holographic renormalisation, a topic to which we now turn.

\subsubsection{Auxiliary field formalism}

To determine the correct counterterms it is useful to consider first the formulation of a 
well-defined boundary value problem with the cut-off radius not taken to infinity.  Since we are dealing
with a higher-curvature theory, it is not enough to fix the metric at the boundary, but also some derivatives 
must be held fixed. A convenient 
and enlightening way to set up the problem, and to choose a combination of derivatives to hold fixed, 
is through the auxiliary field formalism developed in \cite{Bergshoeff:2009hq, Bergshoeff:2009aq, Hohm:2010jc}. 
We shall use a minutely tweaked version, the tweaking being the non-linear
version of the field redefinition used, e.g., in \cite{Grumiller:2009sn}. The field redefinition is such that the
auxiliary field vanishes for the AdS vacuum. 

Running the risk of some redundancy, we present it in some small detail for 
general $\La$ and $\be$. We consider the bulk action
\eq{
S = \int\limits_{M}\extd^4 x \sqrt{-g} \left[(2\La\be + 1)R - (2\La\be + 1)\, 2\La  + \frac{3}{\La}F^{\mu\nu} G_{\mu\nu}
 + \frac{3}{4 \be \La^2}(F^{\mu\nu}F_{\mu\nu} - F^2) + 3F \right]\, , 
}{eq:auxact}
where $F_{\mu\nu}$ is the auxiliary field, $F = g^{\mu\nu}F_{\mu\nu}$ and $G_{\mu\nu} = R_{\mu\nu}- (R/2)g_{\mu\nu}$. 
Varying with respect to $F_{\mu\nu}$, and using the trace of the resulting equation of motion, gives $F_{\mu\nu}$
in terms of the metric:
\eq{
F_{\mu\nu} = \frac{\be\La}{3}\big(-6R_{\mu\nu} + R g_{\mu\nu} + 2\La g_{\mu\nu}\big)\, .
}{eq:F}
From this equation it is easy to see that $F_{\mu\nu} = 0$ if and only if $G_{\mu\nu} + \La g_{\mu\nu} = 0$, 
i.e., the auxiliary field vanishes exactly if the cosmological Einstein tensor vanishes.

Substituting equation \eqref{eq:F} into the action \eqref{eq:auxact} yields after a little algebra
the action \eqref{eq:action} for $\al = -3\be$:
\eq{
S = \int\limits_{M}\extd^4 x\, \sqrt{-g} \left[R - 2\La  - 3\be R^{\mu\nu}R_{\mu\nu}+ \be R^2 \right]\, .
}{}
Note that the action \eqref{eq:auxact} is particularly simple for the critical tuning $2\La\be = - 1$.
In fact, with $\La = -3$ and $\be = 1/6$ we have
\eq{
S^{\rm aux.}_{\rm crit.} = \int\limits_{M}\extd^4 x\, \sqrt{-g} \left[ -F^{\mu\nu} G_{\mu\nu}
 + \frac{1}{2}\big(F^{\mu\nu}F_{\mu\nu} - F^2\big) + 3F \right]\, .
}{}
Apart from the term linear in $F$, this action looks precisely (up to a rescaling of $F_{\mu\nu}$) like the
auxiliary field formulation \cite{Bergshoeff:2009aq} of the pure higher-curvature theory studied by Deser \cite{Deser:2009hb}. 

The boundary value problem is now defined by requiring that the variation of $g_{\mu\nu}$ and $F_{\mu\nu}$
both vanish at the boundary. Thus, the first variation of the on-shell action is allowed to contain boundary terms
multiplying $\de g_{\mu\nu}$ and $\de F_{\mu\nu}$, but no variation of the extrinsic curvature. 
Eliminating such terms requires adding a generalised Gibbons--Hawking term \cite{Hohm:2010jc} which for the critical case 
(and in Gaussian normal coordinates) reads
\eq{
I_{\rm GGH} = \frac{1}{2\kappa^2} \int\limits_{\partial M} \extd^3 x \,\sqrt{-\ga} \, F^{ij}\big(K\ga_{ij} - K_{ij}\big)\, .
}{eq:GGH}
We have now set up a natural-looking boundary value problem for this theory: we keep the metric fixed at the boundary,
and, since non-zero $F_{\mu\nu}$ corresponds to non-Einstein modes, we keep the massive graviton fixed at the boundary.
Note that this is nevertheless a choice --- there are of course other possible boundary conditions.

Fixing this choice now limits the number of additional allowed boundary terms. Most importantly, a boundary term may
not change the boundary value problem. This limits us to use combinations of 
boundary intrinsic metric quantities and the auxiliary field $F_{ij}$. In the next subsection we present a 
set of boundary terms that makes the second variation of the action finite.

\subsubsection{Renormalised second variation}

To regularise the action \eqref{eq:2ndvar}, one needs to find admissible counterterms that cancel all divergences.
Since the computations themselves are not very illuminating, we defer them to appendix \ref{app:2nd}. 
Let us here make some general comments.

First, the generalised Gibbons--Hawking term \eqref{eq:GGH} on its own is very complicated when expanded to 
second order. In particular, there is a non-zero contribution at the order $\sim \log y/y^3$. This is even
more divergent than the terms present in the variation of the bulk action. Furthermore, the first 
variation of $I_{\rm GGH}$ does not vanish. There is, however, a term that remedies these deficiencies. 
The term
\eq{
I_{\hat{F}} = \frac{1}{2\kappa^2} \int\limits_{\partial M} \extd^3 x\, \sqrt{-\gamma}\, \hat{F}  \,,
}{eq:hatF}
where $\hat{F} = \gamma^{ij}F_{ij}$ can be used for this purpose. In fact, the combination
\eq{
I_{\rm GGH} - 2 \, I_{\hat{F}}
}{}
has vanishing first variation, and the second variation starts only at order $\sim 1/y^3$. 
The full result for this term is presented in \eqref{eq:combo}. This term alone, however, far from does the job.
It only cancels one of the problematic terms in \eqref{eq:2ndvar}, and only at the price of adding another
divergent term (at order $\sim \log y/y$). 

To obtain a finite second variation one must add a number of terms involving the field $F_{\mu\nu}$, as well as
the Ricci curvature $R^{(3)}_{ij}$ of the boundary metric $\ga_{ij}$. We have not been able to construct a 
finite action using only such terms however. Instead, to reach our goal requires adding a term of the form 
\eq{
I_{\hat{h}\hat{F}} = \frac{1}{2\kappa^2} \int\limits_{\partial M} \extd^3 x\,\sqrt{-\gamma} \, \hat{h} \, \hat{F} = 
\frac{1}{2\kappa^2} \int\limits_{\partial M} \extd^3 x \,\sqrt{-\gamma} \, \ga^{ij}\big(\ga_{ij} - \eta_{ij}/y^2\big)\hat{F}\, .
}{eq:IhF}
This term is Lorentz invariant at the boundary, but uses not only $\ga_{ij}$ and $F_{ij}$ for its definition, but also
the perturbation $h_{ij}$ explicitly. (Or --- equivalently of course --- it uses the background metric $\eta_{ij}$.)
Although the counterterm \eqref{eq:IhF} does not appear to be generally covariant we find that the boundary stress
tensor is conserved. Thus the theory does not suffer from a diffeomorphism anomaly \cite{AlvarezGaume:1983ig}.
This suggests that the term \eqref{eq:IhF} is actually equivalent to some covariant counterterm to second order in
the perturbation.

The fully regularised action having a finite second variation reads
\eq{
\begin{split}
I_{\rm ren} &= I_{\rm bulk} + I_{\rm GGH} - 2 I_{\hat{F}} \\
 &  - \frac{1}{2\kappa^2} \int\limits_{\partial M} \extd^3 x \, \sqrt{-\gamma} \,\Big[
    \frac{1}{6} F^{ij}F_{ij} -  F^{ij}R^{(3)}_{ij} 
   + \frac{1}{18} F^{ij}\nabla^2 F_{ij} + \frac{5}{18}F^{ij}(D^2F)_{ij} + \hat{h} \, \hat{F} \Big]\, .
\end{split}
}{}
Here, the differential operator $D^2$ is defined in \eqref{eq:D} and computes, up to a factor, the linearised
Ricci tensor around three-dimensional Minkowski space. Expanded in the Fefferman--Graham expansion
\eqref{eq:FGh} the second variation of $I_{\rm ren}$ is 
\eq{
\de^{(2)}I_{\rm ren}|_{\rm EOM} = \frac{1}{2\kappa^2} \int\limits_{\partial M} \extd^3 x\, \sqrt{-\gamma} 
\left[ \frac{3}{2} \de\bup{0}\de\bij{3}
  +\frac{9}{4} \Big(\de\bup{0}\de\gij{3} - \de\hup\de\bij{3} \Big) \right]\, .
}{eq:Itoteom}
Already from this expression it is clear that all two-point correlators involving only Einstein modes vanish:
if all the $b^{(n)}_{ij}$ are zero, so is the second variation. 

\subsection{One-point functions}

The result in Eq.~\eqref{eq:Itoteom} also contains the result for one-point functions around a flat background.
For our operator defined in \eqref{eq:T} we get
\eq{
\langle T_{ij} \rangle =  -\frac{9}{4\kappa^2}\,b^{(3)}_{ij} \,,
 }{eq:<Tij>}
whereas for the operator $t_{ij}$ defined in \eqref{eq:t} we have
\eq{
\langle t_{ij} \rangle = 
  \frac{3}{2\kappa^2}\left( b^{(3)}_{ij} + \frac{3}{2}\,g^{(3)}_{ij}\right)\, .
 }{eq:<tij>}

\section{Two-point correlators}\label{sec:correlators}

The two-point correlators are given by the second variation of the action, or alternatively, by 
the functional derivative of the one-point functions $\langle T_{ij}\rangle$
and $\langle t_{ij}\rangle$ with respect to the sources $\hij$ and $\bij{0}$. Differentiating the
expressions in Eqs.~\eqref{eq:<Tij>} and \eqref{eq:<tij>} for the one-point functions according to 
Eqs.~\eqref{eq:T} and \eqref{eq:t} we obtain
\begin{align}
&\langle T_{ij}(x)\, T_{kl}(x')\rangle = 0 \label{eq:corrEE} \,,\\
&\langle T_{ij}(x)\, t_{kl}(x')\rangle = \frac{9 i}{2 \kappa^2}\frac{\de b^{(3)}_{ij}(x)}{\de b_{(0)}^{kl}(x')} = 
-\frac{9 i}{2 \kappa^2}\frac{\de g^{(3)}_{kl}(x')}{\de h_{(0)}^{ij}(x)} \label{eq:corrlogE} \,,\\
&\langle t_{ij}(x)\, t_{kl}(x')\rangle = -\frac{3i}{\kappa^2}\left(\frac{\de b^{(3)}_{ij}(x)}{\de b_{(0)}^{kl}(x')} + 
\frac{3}{2}\frac{\de g^{(3)}_{ij}(x)}{\de b_{(0)}^{kl}(x')}\right)  \,.  \label{eq:corrloglog}
 \end{align}
Note that there are two ways to compute the $\langle T_{ij}\,t_{kl}\rangle$ correlator --- either by 
differentiating $\langle T_{ij} \rangle$ with respect to $b_{(0)}^{kl}$, or by 
differentiating $\langle t_{kl} \rangle$ with respect to $h_{(0)}^{ij}$. In obtaining the expressions for the correlators we used the fact that
\eq{
\frac{\de b^{(3)}_{ij}}{\de h_{(0)}^{kl}} = 0 \, ,
}{}
meaning that $h^{(0)}_{ij}$ does not source any modes with logarithmic behaviour. We shall see explicitly that this
is true when studying the modes below.

As remarked earlier, correlators including only Einstein modes are identically zero. We immediately
note that any different --- log or otherwise --- behaviour of the log-log correlator with respect
to the log-Einstein correlator stems from the $\gij{3}$ mode, as can be seen from Eq.~\eqref{eq:corrloglog}. 
A logarithmic mode is only defined up to an arbitrary shift of the mode by Einstein modes, and this freedom can be used
to eliminate the first term in Eq.~\eqref{eq:corrloglog}. Thus, 
the non-trivial information in Eq.~\eqref{eq:corrloglog} comes from the last term:
\eq{
\langle t_{ij}\, t_{kl}\rangle = - \frac{2}{3}\langle T_{ij}\, t_{kl}\rangle -\frac{9i}{2\kappa^2}\frac{\de g^{(3)}_{ij}}{\de b_{(0)}^{kl}} \, .
}{eq:corrloglog2}
Eliminating the first term would amount to redefining
\eq{
t_{ij} \to t_{ij} + \frac{1}{3} T_{ij}\, .
}{}
Thus, in order to compute the correlators, all we need to do is find the functional relations between the
matrices $b^{(0)}_{ij}$, $h^{(0)}_{ij}$, $b^{(3)}_{ij}$ and $g^{(3)}_{ij}$. To achieve this we shall find all
linearised modes in momentum space. This is the task of the next subsection.

\subsection{Modes}

We now aim to determine how the subleading Fefferman--Graham coefficients $\bij{3}$ and $\gij{3}$
functionally depend on $\bij{0}$ and $\hij$. This dependence comes about as a combination of the equations
of motion and the boundary conditions in the interior of AdS. In the global case, the latter consists of
requiring regularity, and in the present case of Poincar\'e patch AdS, we require infalling boundary conditions
at the Poincar\'e horizon. 

In transverse gauge, the linearised equations of motion are rather simple. For the critical case they read
\eq{
(\Box + 2)(\Box + 2) \psi_{\mu\nu} = 0\, .
}{eq:lin}
Here $\Box$ is the wave operator on AdS$_4$.
The solution space consists of Einstein modes $\psi^{\rm E}_{\mu\nu}$ and logarithmic modes
$\psi^{\rm log}_{\mu\nu}$ satisfying
\eq{
(\Box + 2)\psi^{\rm E}_{\mu\nu} = 0 \,,
}{}
and 
\eq{
(\Box + 2)^2\psi^{\rm log}_{\mu\nu} = 0\,, \qquad \mbox{but} \qquad (\Box + 2)\psi^{\rm log}_{\mu\nu} \neq 0\, ,
}{}
respectively. Now, we need the modes expressed in Gaussian normal coordinates, which is not compatible with 
transverse traceless gauge in general. We shall anyhow proceed by solving \eqref{eq:lin} and then transform the
solutions to Gaussian normal coordinates. 

To find the logarithmic, as well as the Einstein solutions, we solve the equation 
\eq{
(\Box + m^2) \psi_{\mu\nu}(m) = 0 \,.
}{eq:massive}
Then the Einstein and logarithmic modes are obtained as
\eq{
\psi^{\rm E}_{\mu\nu} = \psi_{\mu\nu}\big(\sqrt{2}\big)\,, \qquad \mbox{and} \qquad
\psi^{\rm log}_{\mu\nu} = \frac{\partial \psi_{\mu\nu}}{\partial m}\Big|_{m=\sqrt{2}}  \,.
}{eq:modes}
We work in three-dimensional momentum space, making the separation ansatz
\eq{
\psi_{\mu\nu} = e^{-ip \cdot x} \tilde\psi_{\mu\nu}(p_i; y)\, .
}{eq:modeansatz}
Using Lorentz invariance of the background, we can fix a certain Lorentzian covector $p_i$.
If $p_i$ is timelike we choose $p_i = E\,\delta^t_i$ and if it is lightlike we choose
$p_i = E\,\delta^t_i+E\,\delta^{x^1}_i$.

\subsubsection*{Timelike modes}

Let us start with the timelike case. Solving \eqref{eq:massive} with 
$\psi_{\mu\nu}(m) = e^{-iEt} \tilde\psi_{\mu\nu}(p_i; y)$, using the gauge condition $\nabla^\mu \psi_{\mu\nu}(m) =0$,
is fairly straightforward. The most general solution has ten undetermined coefficients. 

An example of a solution where eight of the coefficients have been put to zero is
\eq{
\psi_{\mu\nu} = e^{-iEt} \big[C_1 \,  j_{\nu}(E y) + C_2 \, y_{\nu}( E y)\big]
\begin{pmatrix} 0&0&0&0\\ 0&1&0&0\\0&0&-1&0\\0&0&0&0 \end{pmatrix}_{\mu\nu}\, ,
}{eq:example}
where $\nu = 1/2 (-1 + i\sqrt{-17 +4 m^2})$, $j_\nu$ and $y_\nu$ are spherical Bessel functions
and $C_{1,2}$ are constants. (We shall present all solutions only in their final form.) 

The next step is to require infalling boundary conditions at the Poincar\'e horizon at $y=\infty$.
To achieve this we note that 
\eq{
\left[C_1 \,  j_{\nu}(E y) + C_2 \, y_{\nu}( E y)\right] \sim 
\left[C_1 \, \cos(E y - \frac{\pi}{2} -\frac{\nu \pi}{2}) + C_2 \, \sin(E y - \frac{\pi}{2} -\frac{\nu \pi}{2})\right]\frac{1}{y}\, ,
}{}
from which it is clear that infalling boundary conditions correspond to $C_2 = iC_1$. For the other solutions the reasoning is
identical. In this way the number of undetermined components reduces to five. 

The next step is to construct six Einstein and five logarithmic modes by using \eqref{eq:modes}. 
Four of the Einstein modes are pure gauge, whereas all the logarithmic modes are physical. 
The last step is to go to Gaussian normal coordinates. The mode \eqref{eq:example} is already in this gauge.
For those that are not, it is simple to construct the corresponding gauge transformation. 

Below we present the modes that are the result of these computations.
\vspace*{,3cm}
\\
\textit{Logarithmic modes}\vspace*{,3cm}
\\
The full set of logarithmic modes is:
\begin{align}
\psi^{{\rm log}1}_{ij} &= e^{-iEt} \frac{F_1(Ey)}{y^2}
\begin{pmatrix} 0&1&0\\ 1&0&0\\0&0&0 \end{pmatrix}_{ij}\,, \qquad
&\psi^{{\rm log}2}_{ij} &= e^{-iEt} \frac{F_1(Ey)}{y^2}
\begin{pmatrix} 0&0&1\\ 0&0&0\\1&0&0 \end{pmatrix}_{ij}\,, \\
\psi^{{\rm log}3}_{ij} &= e^{-iEt} \frac{F_4(Ey)}{y^2}
\begin{pmatrix} 0&0&0\\ 0&1&0\\0&0&-1 \end{pmatrix}_{ij}\,, \qquad
&\psi^{{\rm log}4}_{ij} &= \frac{e^{-iEt} F_4(Ey)}{y^2}
\begin{pmatrix} 0&0&0\\ 0&0&1\\0&1&0 \end{pmatrix}_{ij}\,, 
\end{align}
\eq{
\psi^{{\rm log}5}_{ij} = \frac{e^{-iEt}}{y^2} \left( F_2(Ey)
\begin{pmatrix} -1&0&0\\ 0&1&0\\0&0&1 \end{pmatrix}_{ij} + F_3(Ey) 
\begin{pmatrix} -1&0&0\\ 0&0&0\\0&0&0 \end{pmatrix}_{ij} \right)\,.
}{}
The functions $F_i(y)$ are defined by
\begin{align}
 F_1(Ey) &= {\rm Ei}(iEy) - e^{i E y}\,, \qquad \qquad F_2 = {\rm Ei}(iEy) - e^{i E y}\left(\frac{3}{2} - \frac{iEy}{2}\right)\,, \\
  F_3(Ey) &= \Big(\frac{E^2 y^2}{2}-3\Big){\rm Ei}(iEy) + 3 e^{i E y}\,,  \\
F_4(Ey) &= e^{-iEy}\Big(-e^{2iEy}\big[2 + \frac{\pi(i+ Ey)}{2}\big] +(Ey - i)(\pi + i{\rm Ei}(2iEy))  \Big)\, .
\end{align}
It is clear that the first two modes correspond to a vector representation of the little group SO$(2)$, 
that $\psi^{{\rm log}3}_{ij}$ and $\psi^{{\rm log}4}_{ij}$ correspond to a traceless tensor representation
and that $\psi^{{\rm log}5}_{ij}$ corresponds to the scalar representation. It is therefore possible to 
immediately write down the modes for general timelike $p_i$.

To this end we denote by $\eps^{1,2}_i$ an orthonormal basis in the orthogonal complement of $p_i$, and construct from 
them two traceless tensors $M^1_{ij} = \eps_i^{1}\eps_j^{1} - \eps_i^{2}\eps_j^{2}$ and 
$M^2_{ij} = \eps_i^{1}\eps_j^{2} + \eps_j^{1}\eps_i^{2}$
in the same space. Then, if $|p| \equiv \sqrt{-p^2}$, we have 
\begin{align}
\psi^{{\rm log}1,2}_{ij} &= e^{-ip\cdot x} \frac{F_1(|p|y)}{y^2} \Big(p_i \eps^{1,2}_j + p_j \eps^{1,2}_i\Big)\,, \\
\psi^{{\rm log}3,4}_{ij} &= e^{-ip\cdot x} \frac{F_4(|p|y)}{y^2} M^{1,2}_{ij} \,,\\
\psi^{{\rm log}5}_{ij} &= \frac{e^{-ip\cdot x}}{y^2} \left( F_2(|p|y)
\eta_{ij} - F_3(|p|y) \frac{p_i p_j}{|p|^2} \right)\, .
\end{align}
\vspace*{,3cm}
\\
\textit{Einstein modes}\vspace*{,3cm}
\\
Regarding the Einstein modes, only the traceless tensor modes are not pure gauge. Thus, log-modes corresponding
to scalar and vector Einstein excitations should be identified as Proca modes \cite{Bergshoeff:2011ri}.\\
To find the pure gauge modes one simply
makes an ansatz $\psi_{\mu\nu} = \nabla_{(\mu} \xi_{\nu)}$ with $\xi_\nu = e^{-iEt}\tilde\xi_{\nu}(y)$ and enforces
Gaussian normal coordinates. The resulting modes read
\begin{subequations}\label{eq:allE}
\begin{align}
\psi^{{\rm E}1,2}_{ij} &= \frac{e^{-ip\cdot x}}{y^2} (p_i \eps^{1,2}_j + p_j \eps^{1,2}_i) \label{eq:EV}\,, \\
\psi^{{\rm E}3,4}_{ij} &= e^{-i(p\cdot x - |p|y)} \frac{(1-i|p|y)}{y^2} M^{1,2}_{ij}  \label{eq:EM} \,,\\
\psi^{{\rm E}5}_{ij} &= -\frac{e^{-ip\cdot x}}{y^2} \left( 2\eta_{ij} - y^2 \, p_i p_j \right) \label{eq:ES1}\,,\\
\psi^{{\rm E}6}_{ij} &= \frac{e^{-ip\cdot x}}{y^2}  p_i p_j \, . \label{eq:ES2}
\end{align}
\end{subequations}
\\
\textit{Functional relations}\vspace*{,3cm}
\\
To derive how $\bij{3}$ and $\gij{3}$ depend on $\bij{0}$ and $\hij$ we must now construct modes that have
either of these two expansion coefficients vanishing. The Einstein modes already have $\bij{0} = 0$, so they correspond
to varying only $\hij$. To get a mode that has vanishing $\hij$ a linear combination of the 
Einstein and logarithmic modes must be taken. To this end, an expansion around $y=0$ for the functions 
$F_i$ is useful:
\eq{
F_1(|p|y) = \log y + \big[-1 + \ga + \frac{i\pi}{2} + \log |p|\big] + \frac{|p|^2 y^2}{4} + \frac{i|p|^3 y^3}{9} + \ldots
}{eq:F1exp}
\eq{
F_2(|p|y) = \log y + \big[-\frac{3}{2} + \ga + \frac{i\pi}{2} + \log |p|\big] - \frac{i|p|^3 y^3}{18} + \ldots
}{}
\eq{
\begin{split}
F_3(|p|y) = &-3 \log y + \big[3 - 3\ga - \frac{3 i\pi}{2} - 3 \log |p|\big] + \frac{|p|^2 y^2}{2}\log y \\
&+ \frac{|p|^2y^2}{4}\big[-3 + 2\ga + i\pi + 2\log |p|\big]
+ \frac{i|p|^3 y^3}{6} + \ldots
\end{split}
}{}
\eq{
\begin{split}
F_4(|p|y) = & \log y 
+ \big[-2 + \ga - i\pi + \log 2 + \log |p|\big] 
+ \frac{|p|^2 y^2}{2}\log y \\
&+ \frac{|p|^2y^2}{2}\big[\ga - i\pi + \log 2 + \log |p|\big]
-\frac{i|p|^3 y^3}{3}\log y\\
&- \frac{i|p|^3 y^3}{9}\big[-8 + 3\ga + \log 8 + 3\log |p|\big] + \ldots
\end{split}
}{eq:F4exp}
Consequently we find the following combinations of logarithmic modes and Einstein modes with vanishing $\hij$:
\begin{align}
 \psi_{ij}^{{\rm L}1,2}&=\psi_{ij}^{{\rm log}1,2}+\big(1-\gamma-\frac{i\pi}{2}-\log |p|\big)\,\psi_{ij}^{{\rm E}1,2} \label{eq:psitildeV} \,,\\
 \psi_{ij}^{{\rm L}3,4}&=\psi_{ij}^{{\rm log}3,4} + \big(2-\gamma+i\pi-\log 2-\log |p|\big)\,\psi_{ij}^{{\rm E}3,4} \label{eq:psitildeM} \,,\\
\begin{split}\label{eq:psitildeS}
 \psi_{ij}^{{\rm L}5}&=\psi_{ij}^{{\rm log}5}-\frac{1}{4}\big(3-2\gamma-i\pi-2\log |p|\big)\,\psi_{ij}^{{\rm E}5} \\
              &\qquad\,\,\,\,+\frac{3}{|p|^2}\,\big(1-\gamma-\frac{i\pi}{2}-\log |p|\big)\,\psi_{ij}^{{\rm E}6} \,.
\end{split}
\end{align}

Let us start with the functional derivative
\eq{
\frac{\delta \gij{3}}{\delta h_{(0)}^{kl}}\, .
}{}
This is the quantity of relevance for computing the two-point function of the energy-momentum tensor in Einstein gravity.
Variation of $g^{(3)}_{ij}$ with respect to $h_{(0)}^{ij}$, keeping $b^{ij}_{(0)}$ fixed, corresponds to analysing the relation between 
$h_{(0)}^{ij}$ and $\gij{3}$ for the Einstein modes. 

We let the index $I$ run over $1, \ldots 6$ and define the basis $\{e^I\}$ of symmetric matrices as 
\eq{
\{e^I\} = \{h_{(0)}^{{\rm E}I}\}
\, ,
}{}
where the subscript $h_{(0)}^{{\rm E}I}$ denotes the leading Fefferman--Graham coefficient of the corresponding Einstein mode 
$\tilde{\psi}^{{\rm E}I}$ from\footnote{We actually take the Fefferman--Graham coefficient of the Fourier transform, i.e., we leave the
exponential factor out. To avoid clutter, we do not put a tilde on $h_{ij}^{(0)}$ and $g_{ij}^{(n)}$. Whenever there is risk of confusion,
we shall write out the argument --- $x$ or $p$ --- explicitly.} 
Eq.~\eqref{eq:allE}. Similarly, by $g_{(3)}^{{\rm E}I}$ below, we shall mean the $\cO(y)$-term in the Fefferman--Graham expansion of the 
corresponding Einstein mode. Explicitly we have
\begin{align}
&e^1_{ij}= p_i \eps_j^{1} + p_j \eps_i^{1}\,, &e_{ij}^2&= p_i \eps_j^{2} + p_j \eps_i^{2}\,,\\
&e^3_{ij}= \eps_i^{1}\eps_j^{1} - \eps_i^{2}\eps_j^{2}\,, &e_{ij}^4&= \eps_i^{1}\eps_j^{2} + \eps_i^{2}\eps_j^{1} \label{basis34}\,,\\
&e_{ij}^5=  - 2\eta_{ij}\,, &e_{ij}^6&= p_i p_j\, .
\end{align}
The only modes having a non-zero $g^{(3)}_{ij}$ are the traceless tensor modes \eqref{eq:EM}.
Explicitly:
\eq{
g^{{\rm E}3,4}_{(3)ij} =  \frac{i|p|^3}{3} e^{3,4}_{ij}\,, \qquad \qquad g^{{\rm E}\neq3,4}_{(3)ij} = 0\, . 
}{eq:g3}
For a general Einstein mode $h_{ij} = \sum_I A_I \tilde{\psi}^{{\rm E}I}_{ij}$ we therefore have 
\eq{
h^{(0)}_{ij} = \sum_I A_I e^I_{ij}\,, \qquad 
g^{(3)}_{ij} = \frac{i|p|^3}{3}\left( A_3  e^3_{ij} + A_4  e^4_{ij}\right)\, .
}{eq:coeffs}
To compute the desired functional relation, we only have left to compute how $A_3$ and $A_4$ depend on $h^{(0)}_{ij}$, 
and to achieve this we need to invert the first relation in \eqref{eq:coeffs}.

Fortunately this is simply done. Noting that 
\eq{
e^{ij}_{K} e^I_{ij} = 2\delta_K^{I}\,, \qquad K = 3,4 \quad\mbox{and} \quad I = 1, \ldots 6\,,
}{eq:invert1}
we have
\eq{
A_K = \frac{1}{2} e^{ij}_{K} h^{(0)}_{ij}\,, \qquad K = 3,4\,.
}{}
Inserting this into the second relation in \eqref{eq:coeffs} we have 
\eq{
g^{(3)}_{ij} = \frac{i|p|^3}{6}\left(e^{3}_{kl}   e^{3}_{ij} +
e^{4}_{kl}e^{4}_{ij}\right)h_{(0)}^{kl}\, ,
}{}
and therefore
\eq{
\frac{\delta g^{(3)}_{ij}}{\delta h_{(0)}^{kl}} = \frac{i|p|^3}{6}\left(e^{3}_{kl} e^{3}_{ij} +
e^{4}_{kl}e^{4}_{ij}\right)\, .
}{eq:funcder}
From the present form it is not clear that this expression is independent of the explicit choice of the
polarization vectors $\eps^{1,2}$, but it is simple to show that this is the case. In fact, defining the matrix
\eq{
\Theta_{ij}(p) = \eta_{ij} p^2 - p_i p_j \, ,
}{eq:momtheta}
and using that
\eq{
\eta_{ij} = \frac{p_i p_j}{p^2} + \eps^1_i \eps^1_j + \eps^2_i\eps^2_j  \,,
}{}
it is straightforward to show that
\eq{
\frac{\delta g^{(3)}_{ij}}{\delta h_{(0)}^{kl}} = 
\frac{i}{6|p|}\big(\Theta_{ik}\Theta_{jl} + \Theta_{il}\Theta_{jk} - \Theta_{ij}\Theta_{kl} \big)\, .
}{eq:momcorrTt}
This expression is identical to the Fourier transform of the two-point function of the stress tensor in a 
three-dimensional conformal field theory \cite{Osborn:1993cr}. (See, e.g., Eqs.~(105) and (108) in Ref.~\cite{Coriano:2012wp}.)

Let us now turn to the functional derivative $\delta \bij{3}/\delta b_{(0)}^{kl}$, which follows from 
a very similar computation. Since we are differentiating with respect to $b_{(0)}^{kl}$ we are interested in the
modes $\tilde{\psi}^{{\rm L}I}$ having vanishing $h^{(0)}_{ij}$. We denote the corresponding Fefferman--Graham coefficients
by $b_{(0)ij}^{{\rm L}I}$ and so on. 

Note first that we have
\eq{
b_{(0)ij}^{{\rm L}I} = e^I\,, \quad I=1,\ldots,4\,\,, \qquad b_{(0)ij}^{{\rm L}5} = -\frac{1}{2}e^5-\frac{3}{p^2}e^6 
= \eta_{ij} -\frac{3}{p^2}p_i p_j\, .
}{eq:b0e}
For an arbitrary mode $h_{ij} = \sum_I A_I \psi^{{\rm L}I}_{ij}$ we furthermore have\footnote{The source $\bij{0}$
has only five degrees of freedom, because it is traceless by the equations of motion \eqref{b0b3g3traceless}. For similar
reasons $\gij{3}$ and $\bij{3}$ have only two degrees of freedom --- they are transverse and traceless.}
\eq{
b^{(0)}_{ij} = \sum\limits_{I=1}^4 A_I e^I_{ij} + A_5 \big(\eta_{ij} -\frac{3}{p^2}p_i p_j\big) \,,
}{eq:coeffs2}
and
\eq{ 
b^{(3)}_{ij} = -\frac{i|p|^3}{3}\left( A_3  e^3_{ij} + A_4  e^4_{ij}\right)\, .
}{eq:coeffs2.5}
Using exactly the same construction as before, we therefore find
\eq{
\frac{\delta b^{(3)}_{ij}}{\delta b_{(0)}^{kl}} = -\frac{\delta g^{(3)}_{ij}}{\delta h_{(0)}^{kl}} = 
-\frac{i}{6|p|}\big(\Theta_{ik}\Theta_{jl} + \Theta_{il}\Theta_{jk} - \Theta_{ij}\Theta_{kl} \big)\, .
}{}
We have now only left to compute $\delta \gij{3}/\delta b_{(0)}^{kl}$. As noted before, the $\gij{3}$ corresponding to 
non-Einstein modes is not necessarily transverse (but always traceless), and these excitations correspond to several distinct operators in the CFT.
Only the transverse traceless tensor part gives the logarithmic partner of the stress tensor.

Using the York decomposition of a traceless tensor, we can split up the operator $t_{ij}$:
\eq{
t_{ij} = \nabla_i V_j + \nabla_j V_i +  t_{ij}^{\rm TT} + \big(\nabla_i \nabla_j - \frac{1}{3}\eta_{ij} \nabla^2\big)S \,,
}{eq:yorktij}
where $V$ is a transverse covector ($\nabla^i V_i =0$), $t_{ij}^{\rm TT}$ is transverse traceless and $S$ is a scalar operator.
The logarithmic partner of the stress tensor is the operator $t_{ij}^{\rm TT}$.

Correspondingly, we shall split up the (traceless) Fefferman--Graham components into three pieces:
\eq{
h_{ij} = \nabla_i v_j + \nabla_j v_i +  h_{ij}^{\rm TT} + \big(\nabla_i \nabla_j - \frac{1}{3}\eta_{ij} \nabla^2\big)s \,.
}{eq:york}
In Fourier space this translates to
\eq{
\tilde{h}_{ij} = -i\left( p_i \tilde v_j + p_j \tilde v_i\right) + \tilde{h}_{ij}^{\rm TT} -
        \big(p_i p_j - \frac{1}{3}\eta_{ij} p^2\big)\tilde s       \,,
}{}
with 
\eq{
p_i \tilde v^i = 0 \qquad {\rm and} \qquad p^i \tilde h_{ij}^{\rm TT} = 0\, .
}{}
Thus, the transverse vector part of $g^{(3)}_{ij}$ corresponds to the $e^1_{ij}$ and $e^2_{ij}$ expansion coefficients,
the transverse tensor part to $e^3_{ij}$ and $e^4_{ij}$, and the scalar part to the $e^5_{ij}$ coefficient.

Let us denote ``transverse vector'' by TV, ``transverse tensor'' by TT and ``scalar'' by S. Then, computing the derivatives
\eq{
\frac{\delta \gij{3}}{\delta b_{(0)}^{kl}}\Big|_{\rm TV}, \qquad \frac{\delta \gij{3}}{\delta b_{(0)}^{kl}}\Big|_{\rm TT} 
\qquad {\rm and} \qquad \frac{\delta \gij{3}}{\delta b_{(0)}^{kl}}\Big|_{\rm S}
}{}
corresponds to letting only $A_{1,2}$, $A_{3,4}$ and $A_5$ be non-zero, respectively. Expanding an arbitrary mode
$h_{ij} = \sum_I A_I \tilde\psi^{{\rm L}I}_{ij}$ to order $y$ produces
\eq{
\begin{split}
g^{(3)}_{ij} = \,&\frac{i|p|^3}{9}\left( A_1  e^1_{ij} + A_2  e^2_{ij}\right)\\
&+\frac{i|p|^3}{3}(C-2\log |p|)\left( A_3  e^3_{ij} + A_4  e^4_{ij}\right)\\
&+\frac{i|p|^3}{18}A_5\Big(\frac{1}{2}e^5 +\frac{3}{p^2}e^6\Big)    \,,
\end{split}
}{eq:coeffs3}
where 
\eq{
C = 14/3 - 2\gamma + \pi i - 2\log 2
}{eq:C} 
is a numerical constant that might in the end be absorbed through redefining the logarithmic mode 
by adding suitable number times the corresponding Einstein mode. Note that the $\log |p|$ term cannot be cancelled in this
way. This is the term responsible for the logarithmic behaviour of the correlators in an LCFT.

To compute the desired variations, we now need to invert Eq.~\eqref{eq:coeffs2} and insert the result in \eqref{eq:coeffs3}.
Again this is straightforward, using that $\{e^1_{ij}, \ldots, e^4_{ij}, b_{(0)ij}^{{\rm L}5}\}$ is an orthogonal basis, and
\eq{\begin{split}
e^{ij}_{1,2} e^{1,2}_{ij} = -2|p|^2\,,  \qquad e^{ij}_{3,4} e^{3,4}_{ij} = 2\,, \qquad b^{(0)ij}_{{\rm L}5}b_{(0)ij}^{{\rm L}5}  = 6\, .
\end{split}
}{eq:invert2}
The result is 
\eq{
\frac{\delta \gij{3}}{\delta b_{(0)}^{kl}} = 
\frac{\delta \gij{3}}{\delta b_{(0)}^{kl}}\Big|_{\rm TV} + \frac{\delta \gij{3}}{\delta b_{(0)}^{kl}}\Big|_{\rm TT} 
+ \frac{\delta \gij{3}}{\delta b_{(0)}^{kl}}\Big|_{\rm S}  \,,
}{}
with 
\begin{subequations}\label{eq:tvtts}
\begin{align}
&\frac{\delta \gij{3}}{\delta b_{(0)}^{kl}}\Big|_{\rm TV} = \frac{2i}{9|p|}p_{(i}\Theta_{j)(l}p_{k)} \label{eq:tv} \,,\\
&\frac{\delta \gij{3}}{\delta b_{(0)}^{kl}}\Big|_{\rm TT} = \big(C - 2\log |p|\big)\frac{\delta \gij{3}}{\delta h_{(0)}^{kl}} \label{eq:tt} \,,\\
&\frac{\delta \gij{3}}{\delta b_{(0)}^{kl}}\Big|_{\rm S} =
\left(p_i p_j - \frac{p^2}{3}\eta_{ij} \right)\left[-\frac{i}{12|p|} \right]\left(p_k p_l - \frac{p^2}{3}\eta_{kl} \right)\, .
\end{align}
\end{subequations}
In \eqref{eq:tv} the symmetrisations are taken over $ij$ and $kl$ and are defined with the usual factor of $1/2$.

\subsubsection*{Spacelike modes}

The above construction of modes captures only the timelike case. There are no obstructions to a very similar
analysis for spacelike modes --- provided we make some obvious changes. For example we will have to choose
different boundary conditions at the Poincar\'e horizon; we demand the modes be regular in the bulk. Finally we end
up with results akin to \eqref{eq:momcorrTt} and \eqref{eq:tt}:
\begin{align}
\frac{\delta g^{(3)}_{ij}}{\delta h_{(0)}^{kl}} =&  \,
\frac{1}{6|p|}\big(\Theta_{ik}\Theta_{jl} + \Theta_{il}\Theta_{jk} - \Theta_{ij}\Theta_{kl} \big)\,, \\
\frac{\delta \gij{3}}{\delta b_{(0)}^{kl}}\Big|_{\rm TT} =&   \,
\frac{C-\log |p|^2}{6|p|}\big(\Theta_{ik}\Theta_{jl} + \Theta_{il}\Theta_{jk} - \Theta_{ij}\Theta_{kl} \big) \,.
\end{align}

\subsubsection*{Lightlike modes}

Similarly we can solve \eqref{eq:massive} with the ansatz $\psi_{\mu\nu}(m)=e^{iE(t+x)}\tilde{\psi}_{\mu\nu}(p_i;y)$
and the gauge condition $\nabla^\mu\psi_{\mu\nu}(m)=0$. However, the modes that we find are all power series in
$y$. Therefore imposing boundary conditions, such as non-singularity of the modes in the bulk, kills one half of all
solutions. Moreover, we kill all possible $\gij{3}$ and $\bij{3}$.
Thus, the integral over the lightlike momenta does not contribute to the correlators. In fact, when actually performing the Fourier
transform in the next subsection we will temporarily pass to Euclidean signature and the complication of having 
lightlike momenta will not play a role. For completeness, we present all lightlike modes in appendix \ref{app:llm}.

\subsection{The correlators}

We have already derived all results needed for computing the correlators in momentum space --- in fact, up to numerical factors, 
the functional derivatives in the above subsection are the momentum space correlators. What remains is  to translate these into
a configuration space form. We shall perform this Fourier transform in Euclidean signature and then continue back to Lorentzian signature.

To illustrate the procedure, a general $h^{(0)}_{ij}(x)$ can be written as [c.f. \eqref{eq:coeffs}]
\eq{
h^{(0)}_{ij}(x) = \frac{1}{(2\pi)^3} \int \extd^3 p \, e^{-ipx} \sum_I A_I(p) e^I_{ij}   \,,
}{eq:ft}
where the factor $1/(2\pi)^3$ is purely conventional. An Einstein mode sourced by this $h^{(0)}_{ij}(x)$ would have
an $\cO(y)$ contribution
\eq{
g^{(3)}_{ij}(x) = \frac{1}{(2\pi)^3} \int \extd^3 p \, e^{-ipx} \frac{i p^3}{3} \big(A_3(p) e^3_{ij} + A_4(p) e^4_{ij}\big)\, . 
}{}
Inverting the Fourier transform and solving for $A_{3,4}$ in \eqref{eq:ft} gives
\eq{
A_{3,4}(p) = \frac{1}{2} e_{3,4}^{kl}(p) \, \int \extd^3 x'  \, e^{ipx'} h^{(0)}_{kl}(x') \, .
}{}
Thus, we have 
\eq{
g^{(3)}_{ij}(x) = \frac{1}{(2\pi)^3} \int \extd^3 p \int \extd^3 x' \, e^{ip(x-x')} \frac{i p^3}{6}
       \big(e^3_{kl} e^3_{ij} + e^4_{kl} e^4_{ij}\big) h_{(0)}^{kl}(x') \, . 
}{}
Comparing with Eq.~\eqref{eq:funcder}, we recognise the Fourier space functional derivative in the integrand. The configuration space
derivative reads
\eq{
\frac{\delta g^{(3)}_{ij}(x)}{\delta h_{(0)}^{kl}(x')} = \frac{1}{(2\pi)^3} \int \extd^3 p  \, e^{ip(x-x')} \frac{\delta g^{(3)}_{ij}}{\delta h_{(0)}^{kl}}(p) \, . 
}{}
Therefore, for the correlators we obtain
\begin{align}
 \langle T_{ij}(x)\,t_{kl}(0) \rangle &= -\frac{1}{(2\pi)^3}\frac{9i}{2\kappa^2} \int \extd^3p\, e^{ipx}
      \frac{\delta g^{(3)}_{ij}}{\delta h_{(0)}^{kl}}(p) \label{xcorrTt}\, , \\
 \langle t_{ij}(x)\,t_{kl}(0) \rangle &= - \frac{2}{3}\langle T_{ij}(x)\, t_{kl}(0)\rangle -\frac{1}{(2\pi)^3} \frac{9i}{2\kappa^2} \int \extd^3p\, e^{ipx}
      \frac{\delta g^{(3)}_{ij}}{\delta b_{(0)}^{kl}}(p) \label{xcorrtt}\, .
\end{align}
To compute these integrals we use the formula
\begin{align}\label{Ftrans}
 \frac{1}{|x|^{2\alpha}} = C(\alpha)\,\int \extd^3p\, \frac{e^{ipx}}{|p|^{2(3/2-\alpha)}}\,, \qquad\qquad
 C(\alpha) =\frac{1}{4^\alpha \pi^{3/2}}\frac{\Gamma\big(\frac{3}{2}-\alpha\big)}{\Gamma(\alpha)} \,,
\end{align}
and its generalisation 
\begin{align}\label{logFtrans}
 -\frac{1}{C(\alpha)}\Big[\frac{\ln |x|^2}{|x|^{2\alpha}}+\frac{C'(\alpha)}{C(\alpha)}\, \frac{1}{|x|^{2\alpha}}\Big]
   =\int \extd^3p\, \frac{e^{ipx}\,\ln |p|^2}{|p|^{2(3/2-\alpha)}} \,,
\end{align}
obtained by differentiation with respect to $\alpha$.
The formulas \eqref{Ftrans} and \eqref{logFtrans} suffice to calculate all configuration space correlators.

As shown in \cite{Osborn:1993cr} the two-point correlator of a spin-two operator $\mathcal{O}_{ij}$ in
three dimensions is given by
\begin{align}\label{EOcorrelator}
 \langle \mathcal{O}_{ij}(x) \mathcal{O}_{kl}(0) \rangle &= \frac{48A}{|x|^6}\,\Big(
       \frac{1}{2}\big(I_{ik}I_{jl}+I_{il}I_{jk}\big)-\frac{1}{3}\eta_{ij}\eta_{kl}\Big) \,,\\
 I_{ij} &= \eta_{ij}+2\frac{x_ix_j}{|x|^2} \,,
\end{align}
where $A$ is a numerical constant. We shall, however, find it more convenient to use the form advocated in \cite{Coriano:2012wp}
\begin{align} \label{dcorrcft}
 \langle \mathcal{O}_{ij}(x) \mathcal{O}_{kl}(0) \rangle &= A \, \hat{\Delta}_{ij,kl} \,\frac{1}{|x|^2}\,,
\end{align}
where
\begin{align}
 \hat{\Delta}_{ij,kl} &=\frac{1}{2}\big(\hat{\Theta}_{ik}\hat{\Theta}_{jl}+\hat{\Theta}_{il}\hat{\Theta}_{jk}\big)
     -\frac{1}{2}\hat{\Theta}_{ij}\hat{\Theta}_{kl} \,, \\
 \hat{\Theta}_{ij} &= \partial_i\partial_j-\eta_{ij}\Box \label{eq:xtheta}\,.
\end{align}
Note that $\hat{\Theta}_{ij}$ defined in \eqref{eq:xtheta}, is the Fourier transform of $\Theta_{ij}$ defined in
\eqref{eq:momtheta}, used to express the correlators in momentum space. Therefore, performing the Fourier transform,
keeping $\hat{\Delta}_{ij,kl}$, is quite trivial and we only need \eqref{Ftrans} and \eqref{logFtrans} for $\alpha=1$. For the correlator
\eqref{xcorrTt} we get
\begin{align}\label{eq:Ttfinal}	
 \langle T_{ij}(x)\,t_{kl}(0) \rangle = A \, \hat{\Delta}_{ij,kl} \,\frac{1}{|x|^2}\,,
\end{align}
 with
\begin{align}\label{eq:A}
 A =  \frac{1}{(2\pi)^3}\frac{6\pi}{\kappa^2}\,.
\end{align}

The $t_{ij}$ operator contains, as explained before, three different pieces. The transverse traceless
part corresponds to the logarithmic partner of the stress tensor. It is clear from Eqs.~\eqref{eq:tvtts} that the correlators of the 
other two operators --- $V_i$ and $S$ from \eqref{eq:yorktij} --- have the usual form for a spin-one and a spin-zero operator:
\begin{align}
\langle V_i(x) V_j(0) \rangle =& \,A_V \,\hat{\Theta}_{ij} \frac{1}{|x|^2} \label{eq:V&S} \,, \\
\langle S(x) S(0) \rangle =& \,A_S \,\frac{1}{|x|^2}\label{eq:V&S2}    \,,
\end{align}
where 
\eq{
A_V = \frac{1}{(2\pi)^3}\frac{\pi}{\kappa^2} \qquad \rm{and} \qquad A_S = -\frac{1}{(2\pi)^3}\frac{3\pi}{2\kappa^2} \, .
}{eq:AVAS}
Note that the two-point correlator of the spin-zero operator is negative.
This is another manifestation of the non-unitarity of the theory and means
that the theory would be non-unitary even if the logarithmic partner of the stress tensor would be absent.

Evaluation of the transverse traceless part of \eqref{xcorrtt} yields, \textit{via} Eq.~\eqref{eq:tt},  
%\begin{align} \label{dcorrlcft}
 %\langle t^{\rm TT}_{ij}(x) t^{\rm TT}_{kl}(0) \rangle &= A \, \hat{\Delta}_{ij,kl} \,\frac{\log|x|^2}{|x|^2}\,.
%\end{align}
\begin{align}\label{eq:ttfinal}
   \langle t^{\rm TT}_{ij}(x)\,t^{\rm TT}_{kl}(0) \rangle &= A\, \hat{\Delta}_{ij,kl} \,\frac{\log|x|^2+C+2\gamma - 2/3}{|x|^2} \,,
\end{align}
generalising \eqref{dcorrcft}.

The ambiguity of the log-mode with respect to addition of Einstein modes is evident by the linearity
of the operator $\hat{\Delta}_{ij,kl}$.  Thus, we can again use the freedom of redefining $t^{\rm TT}_{ij}$ 
to get rid of the factor $C+2\gamma-2/3$ via, $t^{\rm TT}_{ij}\to t^{\rm TT}_{ij}-(C/2+\gamma-1/3)T_{ij}$.

We have now found all two-point functions of four-dimensional critical gravity.
The two correlators \eqref{eq:Ttfinal} and \eqref{eq:ttfinal} together with the result \eqref{eq:A} for the quantity $A$
constitute the main quantitative results of this paper. 

\section{Conclusions}\label{seq:concl}

This work is part of an effort to increase our understanding of higher-curvature gravity in 
four dimensions. In particular, we study the critical tuning of the coupling constants in order to determine 
to what extent the AdS/LCFT duality, discovered in three dimensions, extends to the four-dimensional case. 
As a first step in this direction we computed the one- and two-point functions for the critical theory. 

There are four operators in the boundary theory, which are categorised by their sources.
The stress tensor $T_{ij}$ is sourced by $h^{(0)}_{ij}$ and its one-point function is transverse and traceless
and given by
\eq{
\langle T_{ij} \rangle = 
  -\frac{9}{4\kappa^2}\,b^{(3)}_{ij}\, .
 }{eq:<Tij>concl}
The operators $t_{ij}^{\rm TT}$, $V_i$ and $S$ are sourced by 
the corresponding components of $b^{(0)}_{ij}$, and the one-point functions are 
\begin{align}
\langle t_{ij}^{\rm TT}\rangle &=  \frac{3}{2\kappa^2}\left( b^{(3)}_{ij} + \frac{3}{2}g^{(3),{\rm TT}}_{ij}\right) \,,\\
\langle V_i \rangle &= \frac{9}{4\kappa^2}\,V^{(3)}_{i} \,,\\
\langle S \rangle &= \frac{9}{4\kappa^2}\,S^{(3)}  \,,
\end{align}
where $g^{(3),{\rm TT}}_{ij}$, $V_i^{(3)}$ and $S^{(3)}$ are defined by the York decomposition of $g^{(3)}_{ij}$:
 \eq{
 g^{(3)}_{ij} = \nabla_i V_j^{(3)} +  \nabla_j V_i^{(3)} + g^{(3),{\rm TT}}_{ij} +
              \big(\nabla_i \nabla_j - \frac{1}{3}\eta_{ij} \nabla^2\big)S^{(3)}\,.
 }{}
 Note that the one-point function of the vector operator satisfies a Ward identity of the form $\nabla^i \langle V_i \rangle = 0$.
 
 The fact that the stress tensor vanishes for all Einstein solutions is consistent with the result that the
 mass and entropy of black holes in the theory vanish \cite{Lu:2011zk}.
 
The non-trivial two-point correlators all match the expectations for a logarithmic CFT,
$T_{ij}$ and $t^{\rm TT}_{ij}$ forming a rank-two logarithmic pair. Explicitly, 
 \begin{subequations}\label{eq:rank2}\begin{align}
 \langle T_{ij}(x)\,T_{kl}(0) \rangle &=0 \,, \\
 \langle T_{ij}(x) \, t^{\rm TT}_{kl}(0) \rangle &= \frac{1}{(2\pi)^3}\frac{6\pi}{\kappa^2} \,\hat{\Delta}_{ij,kl}\frac{1}{|x|^2} \,, \\
 \langle t^{\rm TT}_{ij}(x) \, t^{\rm TT}_{kl}(0) \rangle &= \frac{1}{(2\pi)^3}\frac{6\pi}{\kappa^2}
          \,\hat{\Delta}_{ij,kl}\frac{\log(|x|^2 m^2)}{|x|^2} \label{loglog} \,,
\end{align}\end{subequations}
where we parameterised the freedom to add a multiple of an Einstein mode to the log-mode by a fiducial mass scale $m$
in \eqref{loglog}. The two-point functions of $V_i$ and $S$ have the form expected for ordinary spin-one and spin-zero operators
\eqref{eq:V&S}--\eqref{eq:AVAS}, and all ``mixed'' correlators vanish.
The relatively simple form of the log-log correlator \eqref{loglog} suggests a natural generalisation
to arbitrary dimensions:
\eq{
 \langle t^{\rm TT}_{ij}(x) \, t^{\rm TT}_{kl}(0) \rangle \propto
     \,\hat{\Delta}_{ij,kl}^{(d)}\frac{\log|x|^2}{|x|^{2d-4}}  \,,
}{}
where
\eq{
      \hat{\Delta}_{ij,kl}^{(d)} =\frac{1}{2}\big(\hat{\Theta}_{ik}\hat{\Theta}_{jl}+\hat{\Theta}_{il}\hat{\Theta}_{jk}\big)
     -\frac{1}{d-1}\hat{\Theta}_{ik}\hat{\Theta}_{jl} \,.
}{}
To explicitly derive this result, and to compute the constant of proportionality, could be a worthwhile exercise. 

The result \eqref{eq:rank2} demonstrates that a lot of the story from three dimensions is repeated in the present case.
There are, however, also differences. In three dimensions, the logarithmic graviton corresponds to two degrees of freedom, 
each playing the role of one chiral component of the logarithmic partner of the stress tensor.
In the present case, this part is played by the transverse and traceless tensor modes. But now there are three additional 
degrees of freedom, parameterised by $V_i$ and $S$ above. These were called Proca modes in \cite{Bergshoeff:2011ri}, 
and their normalisable representatives do not have logarithmic fall-off at the conformal boundary ($b^{(3)}_{ij} = 0$).
The spin-zero operator gives rise to a negative two-point function, see \eqref{eq:V&S2} together with \eqref{eq:AVAS}.
This is another indication of the non-unitarity of the theory.
% Note that the two-point correlator of the scalar excitation, see \eqref{eq:V&S2} and \eqref{eq:AVAS}, is negative.
% This is an other manifestation of the non-unitarity of the theory and means
% that the theory would be non-unitary even if the logarithmic partner of the stress tensor would be absent.

The Proca modes have zero energy by \eqref{eq:<Tij>concl}. It would be very interesting to know whether these modes are subject to
a linearisation instability akin to that found in \cite{Maloney:2009ck}. If they do show a linearisation instability, then
all logarithmic modes can possibly be truncated by imposing boundary conditions, leaving a theory propagating only massless
(and energyless!) spin-two gravitons. Otherwise, such a truncation is impossible. We note that there are 
other consistency requirements for a truncation to work. The boundary conditions must be consistent and the higher-point
functions between truncated and untruncated modes must vanish. See, e.g., the 
discussions in \cite{Skenderis:2009kd, Grumiller:2009mw}.

Another interesting extension of the present work would be to extend the analysis to the higher-derivative gravity models 
recently presented in \cite{Nutma:2012ss}. Degeneration of multiple massive modes with each other and/or the Einstein
modes might lead to gravity duals for higher-rank LCFTs in arbitrary dimensions. Such an extension would then involve the 
calculation of two-point functions in higher-rank Jordan cells. 

Furthermore, while critical gravities are non-unitary \cite{Porrati:2011ku} (as are all duals of LCFTs), critical tunings
of higher-derivative models that lead to odd-rank LCFTs, might open up the interesting possibility of unitary truncations
\cite{Bergshoeff:2012sc}. First steps of such a venture in the gravity context are taken in \cite{Bergshoeff:2012ev}. 

Last, but not least, it would of course be very rewarding to find a suitable condensed matter application for higher-dimensional
LCFT, and to construct a phenomenological holographic model of it, using the model explored in this paper as canvas.

\section*{Acknowledgements}

Advice and support from Mohsen Alishahiha and Daniel Grumiller has been essential throughout this work. 
We furthermore thank Hamid Afshar, Eric Bergshoeff, Ricardo Caldeira Costa, 
Sabine Ertl, Michael Gary, Jan Rosseel, Kostas Skenderis, Marika \makebox{Taylor} and Erik Tonni for discussions.
NJ and TZ were supported by the START project Y435-N16 of the Austrian Science Fund (FWF) 
and by the FWF project P21927-N16. AN is supported by the Ministry of Science, Research
and Technology in Iran, Iran National Science Foundation (INSF) and by Nederlandse Organisatie voor Wetenschappelijk Onderzoek
via a VICI grant of Kostas Skenderis. AN wishes to thank ITFA for hospitality at 
various stages of this project and TZ thanks the Erwin Schr\"odinger International Institute for Mathematical Physics for their 
kind hospitality during the ``Workshop on Higher Spin Gravity" where the project was finalised. 

Finally, NJ would like to express his heartfelt thanks to all colleagues and co-authors through the years. 

\appendix

\section{Global coordinates: Fefferman--Graham expansion and first variation}\label{FG}

\subsection{First variation of the action}

In the Gaussian normal coordinates of Eq.~\eqref{eq:FG1} we have
\eq{
K_{ij}= \frac{1}{2}\partial_{\rho} \ga_{ij}, \qquad \Ga^i_{j\rho} = K^i_j\, ,  
\qquad \Ga^\rho_{ij} = -K_{ij}\, .
}{}
The matrix $\ga_{ij}$ is assumed to have the expansion \eqref{eq:FG2} with the leading contributions fixed.

In Gaussian normal coordinates and after partially integrating the variations of the Christoffel symbols, the boundary term coming from the first variation of the bulk action becomes
\eq{
\begin{split}
\de I_{\rm bulk} |_{\rm EOM} = \frac{1}{2\ka^2}&\int_{\partial M} \extd^3x \,\sqrt{-\ga} \, J^{\rho}
 = \int_{\partial M} \extd^3 x\, \sqrt{-\ga}\,\Big( -(A^{ij} + A^{\rho\rho}\ga^{ij})\de K_{ij} \\
&+ \left[\frac{1}{2}\nabla_{\rho}(A^{ij} + A^{\rho\rho}\ga^{ij}) + \nabla_k A^{\rho k}\ga^{ij} - \nabla^i A^{\rho j} + A^{\rho\rho}K^{ij} \right]
\de \ga_{ij}
\Big)\, .
\end{split}
}{eq:1stvar}
For the case of interest $\al = -3\be$, the tensor $E_{\mu\nu}$ of Eq.~\eqref{eq:var2} is traceless, and the trace of the equations of motion thus implies 
$R = 4\La = -12$. Asymptotically this implies the tracelessness of $\be^{(3)}$ and $\ga^{(3)}$:
\eq{
\Tr \be^{(3)} \equiv \ga^{ij}_{(0)}\be^{(3)}_{ij} = \Tr \ga^{(3)} \equiv \ga^{ij}_{(0)}\ga^{(3)}_{ij} = 0\, .
}{eq:FG4}
Below the symbol $\approx$ denotes equalities after these equations have been imposed. A quantity of interest is
the Ricci tensor. We obtain the following expansions for its components:
\eq{
\begin{split}
R_{ij} &= -3\ga_{ij} + \frac{e^{-\rho}}{2} \left[( -3 \rho + 1)\Tr \be^{(3)} + 3 \Tr \ga^{(3)} \right] -\frac{3 e^{-\rho}}{2}\be_{ij}^{(3)}
+ \cO(e^{-2\rho}) \\
&\approx  -3\ga_{ij} -\frac{3 e^{-\rho}}{2}\be_{ij}^{(3)}  + \cO(e^{-2\rho})\, ,
\end{split}
}{eq:FG5}
\eq{
R_{i\rho} = -e^{-\rho} R_{iy} = \cO(\rho e^{-3\rho})\, , 
}{eq:FG6}
\eq{
\begin{split}
R_{\rho\rho} &= e^{-2\rho} R_{yy} = -3 + \frac{1}{2} e^{-3\rho} \left([3\rho-4]\Tr \be^{(3)} -  3 \Tr \ga^{(3)} \right) + \cO(\rho e^{-5\rho})\\
&\approx  -3 + \cO(\rho e^{-5\rho})\, ,
\end{split}
}{eq:FG6.5}
\eq{
R \approx -12 + \cO(e^{-4\rho})\, .
}{eq:FG6.6}
Other useful expressions are (indices are raised by $\ga_{(0)}^{ij} = [\ga_{(0)}^{-1}]^{ij}$)
\eq{
\ga^{ij} = e^{-2\rho} \ga_{(0)}^{ij} - e^{-4\rho} \ga_{(2)}^{ij} + \rho \, e^{-5\rho} \be_{(3)}^{ij} - e^{-5\rho} \ga_{(3)}^{ij} \,,
}{eq:FG7}
\eq{
K_{ij} = e^{2\rho}\gaij0 + \frac12 (\rho -1) e^{-\rho} \beij{3} - \frac12 \, e^{-\rho} \gaij{3} + \ldots 
}{eq:FG8}
\eq{
K^i_j = \de^i_j - \erho{-2}\ga^{i}_{(2)}{}_{j} + \rho \, \erho{-3} \frac{3}{2} \be^{i}_{(3)}{}_{j} - \erho{-3}[\frac{1}{2} \be^{i}_{(3)}{}_{j} +  \frac{3}{2} \ga^{i}_{(3)}{}_{j}] + \ldots
}{eq:FG9}
\eq{
K^{ij} = \erho{-2} \ga^{ij}_{(0)} - 2\erho{-4} \ga^{ij}_{(2)} + \frac{5}{2} \rho \, \erho{-2} \be^{ij}_{(3)}
-\frac12 \, \erho{-5}\, [\be^{ij}_{(3)} + 5\ga^{ij}_{(3)}] 
+ \ldots
}{eq:FG10}

\eq{
\sqrt{-\ga} = \sqrt{\ga^{(0)}} \left(\erho3 - e^\rho \right) + \ldots
}{eq:FG11}
We use these expressions to obtain the following expansions
\eq{
A^{ij} \approx (1-6\be)\ga^{ij} + 9 \be \, \erho{-5}  \be^{ij}_{(3)} + \ldots
}{eq:FG12}
\eq{
A_{\rho\rho} = A^{\rho\rho} \approx (1-6\be) + \cO(e^{-4\rho})\, ,
}{eq:FG13}
\eq{
\nabla_i A^{\rho j} \approx -9\be \be^j_{(3)}{}_{i}\, \erho{-3} + \cO(\erho{-4}) \, ,
}{eq:FG14}
\eq{
\nabla_k A^{\rho k} \approx \cO(\erho{-4})\, ,
}{eq:FG15}
\eq{
\nabla_\rho A^{\rho \rho} \approx \cO(\erho{-4})\, ,
}{eq:FG16}
\eq{
\nabla_\rho A^{ij} \approx -27\,\be\, \be^{ij}_{(3)} \erho{-5} + \ldots 
}{eq:FG17}
Putting these expressions together yields for the variation of the on-shell action
\eq{
\begin{split}
\de I_{\rm bulk} |_{\rm EOM} &= \frac{1}{2\ka^2}\int_{\partial M} \extd^3x\,\sqrt{-\ga} \, J^{\rho} = \\
&= \frac{1}{2\ka^2}\int_{\partial M} \extd^3x\,\sqrt{-\ga} \left([1-6\be](K^{ij}\de \ga_{ij} - 2 \ga^{ij}\de K_{ij} )  - \frac{27\be}{2} \erho{-5} \be^{ij}_{(3)} \de \ga_{ij}\right), 
\end{split}
}{eq:varon}
which is the expression quoted in the main text. 

For $\be=1/6$ we do not need any holographic counterterms to get a well-defined variational principle.
This is similar to the three-dimensional case, where we did not need any counterterms for the logarithmic
point of NMG \cite{Hohm:2010jc,Alishahiha:2010bw}. The stress tensor can be directly read off from the above
equation and is proportional to $\be_{(3)}^{ij}$.

If $\be\neq 1/6$ (in which case $\be^{ij}_{(3)}$ vanishes by virtue of the equations of motion) we need a holographic
counterterm. It turns out that the same counterterms as for pure four-dimensional gravity \cite{Balasubramanian:1999re,deHaro:2000xn}
multiplied with a numerical factor does the job.

After adding this counterterm, the action reads
\eq{
\begin{split}
I_{\rm ren} &= \frac{1}{2\ka^2} \int_M \extd^4 x \, \sqrt{-g} \,  \Big[R - 2\La + \be(R^2- 3R^{\mu\nu}R_{\mu\nu})\Big] \\
	&  - \frac{1-6\be}{2\ka^2} \int_{\partial M} \extd^3 x \, \sqrt{-\ga} \, \Big(4-2K+R[\ga] \Big) \, ,
\end{split}
}{eq:Iren}
where $R[\ga]$ is the contracted Ricci tensor of the induced metric $\ga_{ij}$ on the boundary.
Expressed in the components of the full Ricci tensor this quantity is given by
\eq{
R[\ga]^{ij} = R^{ij} + \partial_\rho K^{ij} + KK^{ij} + 2 K^{il} K_l^j\, . 
}{}
Using the expansions \eqref{eq:FG5}--\eqref{eq:FG10} it is now straightforward to obtain
\eq{
R[\ga]^{ij} = -\ga_{(2)}^{ij} e^{-4\rho} + 2\ga^{ij}_{(0)}e^{-4\rho} \, . 
}{}
Armed with this equation we derive the variation of the boundary term:
\eq{
\begin{split}
\de \Big(\sqrt{-\ga}(4 - 2K + R[\ga]) \Big) &= \sqrt{-\ga}\Big(-2\ga^{ij} \de K_{ij} 
+ [2K^{ij} -\ga^{ij} + \ga^{ij}_{(2)}e^{-4\rho} ]\de \ga_{ij} \Big)\\ &+ \cO(\rho e^{-\rho})\, .
\end{split}
}{eq:varbdry}
This allows us to determine the variation of the full action \eqref{eq:Iren}. Adding the contributions from
Eqs.~\eqref{eq:varon} and \eqref{eq:varbdry} produces ($\be^{(3)}_{ij} =  0$)
\eq{
\de I_{\rm ren} |_{\rm EOM} =  \frac{1}{2\ka^2} \int_{\partial M} 
\extd^3x \, \sqrt{-\ga^{(0)}} \, \frac{3}{2}\, (1-6\be) \, \ga^{ij}_{(3)} \, \de \ga^{(0)}_{ij}\, .
}{}
This shows that the stress tensor in this case is proportional to $\ga^{ij}_{(3)}$. As noted in the main text, 
because we ignore other fall-off behaviours than present in \eqref{eq:FG2}, this result does not contain
contributions from massive gravitons, but only from massless gravitons and black holes.

\subsection{Asymptotic equations of motion}

Let us now turn to the asymptotic equations of motion.
By taking the trace of the EOMs we already established
tracelessness of $\ga^{(3)}_{ij}$ and $\be^{(3)}_{ij}$. To show that the stress tensors for the critical case
($\be=1/6$) and the non-critical case ($\be\neq1/6$) are conserved we need to take a look at the $i\rho$-components
of the EOMs. For $\be=1/6$ we use the ansatz \eqref{eq:FG2} and plug it into the equations of motion. 
Using the simplification that follows from that $R$ is constant and throwing away terms proportional to the
traces of $\ga^{(3)}_{ij}$ and $\be^{(3)}_{ij}$ we obtain
\eq{
{\rm EOM}_{i\rho} =  \frac{9}{4} \, \nabla_{(0)}^k \be_{ki}^{(3)} e^{-3\rho} + \cO(\rho e^{-4\rho}) \, .
}{eq:divbeta}
Here $ \nabla_{(0)}$ is the covariant derivative with respect to $\ga^{(0)}_{ij}$. This proves the conservation law for the critical case.

In the non-critical case $\be\neq1/6$ we do not have logarithmic modes and therefore we have to omit the
$\be^{(3)}_{ij}$ term in the expansion \eqref{eq:FG2}. Again the $i\rho$-components of the EOMs yield
\eq{
{\rm EOM}_{i\rho} = \frac{3}{2} (1 -6\be ) \, \nabla_{(0)}^k \ga_{ki}^{(3)}e^{-3\rho} + \cO(\rho e^{-4\rho}) \, ,
}{eq:divgamma}
up to trace terms. Thus, the stress tensors \eqref{eq:tijcrit} and \eqref{eq:tij} for the action \eqref{eq:action} are 
traceless and conserved for any value of $\be$.

\section{Poincar\'e coordinates: Fefferman--Graham expansion and second variation} \label{app:2nd}

We expand around Poincar\'e patch AdS:
\eq{
\extd s^2 = \frac{\extd y^2}{y^2} + \ga_{ji}\extd x^i \extd x^j =
 \frac{\extd y^2 + \eta_{ij}\extd x^i \extd x^j}{y^2} + h_{ij}\extd x^i \extd x^j \,,
}{}
where $\eta_{ij}$ is the flat 3d Minkowski metric, and $h_{ij}$ has the expansion
\eq{
h_{ij} = b^{(0)}_{ij}\frac{\log y}{y^2} + h^{(0)}_{ij} \frac{1}{y^2} + b^{(2)}_{ij}\log y +  g^{(2)}_{ij}
+ b^{(3)}_{ij}\, y \, \log y +  g^{(3)}_{ij}\, y + \ldots 
}{}
near $y=0$. Our goal is to compute the variation of the terms multiplying the variations $\delta \gamma_{ij}$ and $\de K_{ij}$ in 
\eqref{eq:1stvar}. To give them names, let us define 
\eq{
\de I_{\rm bulk} |_{\rm EOM} = \frac{1}{2\ka^2} \int_{\partial M} \extd^3x\,\sqrt{-\ga} \, J^{\rho}
 = \int_{\partial M} \extd^3 x\, \sqrt{-\ga}\,\Big( R_1^{ij}\de K_{ij} + R_2^{ij}\de \ga_{ij}
\Big)\, .
}{}
We have computed $R_1^{ij}$ and $R_2^{ij}$ only on-shell in \eqref{eq:1stvar}, but this is all we need. In fact, to compute the correlators, we
put in variations that satisfy the linearised equations of motion. Therefore, it is enough to vary the on-shell quantities, and
we may, along the way, use on-shell relations between the Fefferman--Graham coefficients.

For the purpose of expanding $R_1^{ij}$ and $R_2^{ij}$, and to expand the equations of motion, 
we need the Ricci tensor to the first order in $h_{ij}$. It is most simply obtained using some
computer algebra package (we used \textit{GRTensorII} \cite{grtensor}) and it reads
\begin{align}
 R_{ij} =&-3\ga_{ij} - \frac{y^2}{2}(D^2 \ga)_{ij} + \frac{3}{2 y^2} b^{(0)}_{ij} + (\log y -1/2)b^{(2)}_{ij}
 + g^{(2)}_{ij} - \frac{3y}{2}b^{(3)}_{ij} \nonumber \\
 &+ \Big[ \frac{1}{2y^2} \Tr b^{(0)} + (\log y + 1/2) \Tr b^{(2)} + \Tr g^{(2)} \\
 & \qquad + \frac{y}{2}\left(1 + 3 \log y \right)\Tr b^{(3)} + \frac{3y}{2} \Tr g^{(3)}\Big] \eta_{ij} + \cO(y^2 \log y) + \cO(h^2) \,,\nonumber\\
 R_{iy} =& \frac{1}{2y}\left(\nabla_k b^k_{(0)i}-\nabla_i \Tr b^{(0)} \right) 
 + (\log y + 1/2)\, y\, \left(\nabla_k b^k_{(2)i}-\nabla_i \Tr b^{(2)} \right)\\
 &+ \left(\nabla_k g^k_{(2)i}-\nabla_i \Tr g^{(2)} \right) \,,\nonumber\\
 R_{yy} =& \frac{1}{y^2}\left(-3 + \Tr b^{(0)} - y^2 \Tr b^{(2)} - \frac{y^3}{2}(3\log y + 4)\Tr b^{(3)}-\frac{3y^3}{2}\Tr g^{(3)} \right)
 \, .
\end{align}
In these equations, and always when it comes to the quantities $g^{(n)}_{ij}$ and $b^{(n)}_{ij}$, the trace is taken with respect to
the metric $\eta_{ij}$, i.e., $\Tr b^{(n)} = \eta^{ij}b^{(0)}_{ij}$. Also, the indices of $g^{(n)}_{ij}$ and $b^{(n)}_{ij}$ are always 
raised by $\eta^{ij}$ and all covariant derivatives are with respect to $\eta_{ij}$, i.e., are just ordinary derivatives. 
We also defined the differential operator $D^2$ as
\eq{
(D^2 g^{(n)})_{ij} = \nabla_i \nabla_j (\Tr g^{(n)}) + \nabla^2 g^{(n)}_{ij} - (\nabla_i \nabla_k g^k_{(n) j} + 
\nabla_j \nabla_k g^k_{(n) i})\, .
}{eq:D}
Note that the operator $D^2$ computes, up to a factor of $-1/2$, the linearised Ricci scalar corresponding to 
a metric perturbation $g^{(n)}_{ij}$ around a flat background. 

\subsection{Linearised equations of motion}

Let us start by expanding the linearised version of ${\rm EOM}_{\mu\nu}$ as
\eq{
{\rm EOM}^{(1)}_{\mu\nu} = \sum\limits_{n=-1}^{\infty} \left[ ({\rm EOM}^{L}_n)_{\mu\nu} y^n \log y 
+ ({\rm EOM}_n)_{\mu\nu} y^n 	\right]
}{}
and the linearised Ricci scalar as
\eq{
R^{(1)} = \sum\limits_{n=0}^{\infty} \left[ R^{L}_n y^n \log y + R_n y^n \right]\, .
}{}
Evaluation of these quantities, which all must vanish, yields the linearised equations of motion.
From the constancy of the Ricci scalar we obtain the following relations ($R_0^L$,  $R_1^L$ and $R_1$ are all 
manifestly zero)
\begin{align}
R_0 =& -4 \Tr b^{(0)} = 0 \,,\\
R^{L}_2 =& \nabla^2 \Tr b^{(0)} - \nabla_ i \nabla_j b^{ij}_{(0)} - 4\Tr b^{(2)}= 0 \,,\\
R_2 =& \nabla^2 \Tr h^{(0)} - \nabla_ i \nabla_j h^{ij}_{(0)} - 4\Tr g^{(2)} = 0\,,\\
R_3^L =& -3 \Tr b^{(3)} = 0\,,\\
R_3 =& -3 \Tr g^{(3)} - 2 \Tr b^{(3)}= 0\,,
\end{align}
from which we deduce
\begin{align}
&\Tr b^{(0)} = \Tr g^{(3)} = \Tr b^{(3)} = 0\,, \label{b0b3g3traceless}\\
&\nabla_ i \nabla_j b^{ij}_{(0)} + 4\Tr b^{(2)} = 0 \,,\\
&\nabla^2 \Tr h^{(0)} - \nabla_ i \nabla_j h^{ij}_{(0)} - 4\Tr g^{(2)} = 0\, .
\end{align}
Using these equations to simplify the expressions coming from $({\rm EOM}^{L}_0)_{ij}$ and 
$({\rm EOM}_0)_{ij}$ yields
\begin{align}
&(D^2 b^{(0)})_{ij} = 2 b^{(2)}_{ij} + 2 \Tr b^{(2)} \eta_{ij} \,,\\
&(D^2 h^{(0)})_{ij} = -\frac{1}{2} \nabla^2 b^{(0)}_{ij} - 3 b^{(2)}_{ij} + 2g^{(2)}_{ij} 
+ \left[2 \Tr g^{(2)} + \Tr b^{(2)}\right]\eta_{ij} \, .
\end{align}

\subsection{Second variation of the action}

Using the equations of motion derived in the last subsection, we now want to compute the action to second order in 
the perturbation and expand it in $y$. Also this is best done using computer algebra.
When we present the result we shall not keep track of the two variations separately since, for computing
correlators, all we need are the symmetrised variations. More explicitly, if, say
\eq{
\de^{(2)}S[\delta_1 \gamma_{ij}, \delta_2 \gamma_{ij}] = \frac{1}{2\kappa^2} \int\limits_{\partial M} \extd^3 x \,\sqrt{-\gamma}\,
   (\de_1 h_{(0)}^{ij}\de_2 g^{(3)}_{ij} + \ldots )\,,
}{eq:full2}
we shall only write
\eq{
\de^{(2)}S = \frac{1}{2\kappa^2} \int\limits_{\partial M} \extd^3 x \,\sqrt{-\gamma} \, ( \de h_{(0)}^{ij} \de g^{(3)}_{ij} + \ldots )\, .
}{}
In this appendix, we shall often even leave the integral sign, the factors of $1/2\kappa^2$ and $\sqrt{-\gamma}$, and the $\de$ out and write
\eq{
\de^{(2)}S =  h_{(0)}^{ij} g^{(3)}_{ij} + \ldots 
}{}
when actually meaning \eqref{eq:full2}, hoping that this does not lead to confusion. 

Using these conventions, and using the equations of motion, the second variation of the bulk action reads
\eq{
\begin{split}
\de^{(2)}I_{\rm bulk}|_{\rm EOM} = &-\frac{3}{4} \bup{0}\bij{0} \frac{1}{y^3} 
+ \left[ -\frac{1}{2}\bup{0}\bij{2} + 2\hup \bij{2} - 2 \bup{0}\gij{2} \right] \frac{1}{y} \nonumber \\
&-\frac{9}{2} \bup{0}\bij{3} \log y - \frac{9}{4}\bup{0}\gij{3} - \frac{9}{4}\hup \bij{3}\, .
\end{split}
}{}

\subsection{Boundary terms}

We now apply the same computational and notational framework to various counterterms. The terms in
question are ($F_{\mu\nu}$ is the auxiliary field and $\hat{F} = \ga^{ij}F_{ij}$)
\begin{align}
I_{\rm GGH} &= \frac{1}{2\kappa^2} \int\limits_{\partial M} \extd^3 x \sqrt{-\gamma} F^{ij}(K\ga_{ij} - K_{ij}) \,,\qquad
&I_{\hat{F}} &= \frac{1}{2\kappa^2} \int\limits_{\partial M} \extd^3 x \sqrt{-\gamma}\, \hat{F} \,, \nonumber \\
I_{FF} &= \frac{1}{2\kappa^2} \int\limits_{\partial M} \extd^3 x \sqrt{-\gamma} \, F^{ij}F_{ij} \,, \qquad
&I_{F R} &= \frac{1}{2\kappa^2} \int\limits_{\partial M} \extd^3 x \sqrt{-\gamma}\, F^{ij}R^{(3)}_{ij} \,, \nonumber \\
I_{F \nabla F} &= \frac{1}{2\kappa^2} \int\limits_{\partial M} \extd^3 x \sqrt{-\gamma}\, F^{ij}\nabla^2 F_{ij} \,, \qquad
&I_{F D F} &= \frac{1}{2\kappa^2} \int\limits_{\partial M} \extd^3 x \sqrt{-\gamma}\, F^{ij}(D^2F)_{ij}\nonumber \,,\\
I_{\hat{h}\hat{F}} &= \frac{1}{2\kappa^2} \int\limits_{\partial M} \extd^3 x \sqrt{-\gamma}\,\ga^{ij}(\ga_{ij} - \eta_{ij}/y^2)\hat{F} \, .&&
\end{align}
The second variations of these quantities are\footnote{The generalised Gibbons--Hawking term $I_{\rm GGH}$ actually has a 
non-vanishing first variation, but a linear combination of $I_{\rm GGH}$ and $I_{\hat{F}}$ has vanishing first variation.
The variation of this combination is displayed in \eqref{eq:combo}.} 
\begin{align}
 \de^{(2)}( I_{\rm GGH} - 2 I_{\hat{F}}) &= 
 \frac{3}{2} \bup{0}\bij{0} \frac{1}{y^3} + 3 \bup{0}\bij{2} \frac{\log y}{y}\nonumber \\
&+ \left[\bup{0}\bij{2} - \hup \bij{2} +\Tr b^{(2)}\Tr h^{(0)} + 4 \bup{0}\gij{2} \right] \frac{1}{y}\label{eq:combo}\\
&+ \frac{9}{2} \bup{0}\bij{3} \log y + \frac{9}{2}\bup{0}\gij{3}  \,,\nonumber
\end{align}
\eq{
\de^{(2)}I_{FF} = 
 \frac{9}{2} \bup{0}\bij{0} \frac{1}{y^3} 
 + \left[-3\bup{0}\bij{2} - 6\hup \bij{2} + 6\Tr b^{(2)}\Tr h^{(0)} + 6 \bup{0}\gij{2} \right] \frac{1}{y} - 9\bup{0}\bij{3} \,,
}{}
\eq{
\de^{(2)}I_{F R} = 
 - 3 \bup{0}\bij{2} \frac{\log y}{y} + \left[-3\hup \bij{2} + 3\Tr b^{(2)}\Tr h^{(0)}  \right] \frac{1}{y}  \,,
}{}
\eq{
\de^{(2)}I_{F \nabla F} = 
\left[-27\bup{0}\bij{2}  - 18 \hup \bij{2} + 18 \Tr b^{(2)}\Tr h^{(0)} + 18 \bup{0}\gij{2} \right] \frac{1}{y}  \,,
}{}
\eq{
\de^{(2)}I_{F D F} = 9\bup{0}\bij{2}\frac{1}{y}  \,,
}{}
\eq{
\de^{(2)}I_{\hat{h}\hat{F}} = 2 \Tr b^{(2)}\Tr h^{(0)} \frac{1}{y}  \,.
}{}
Combining these results and defining
\eq{
I_{\rm tot} = I_{\rm bulk} + I_{\rm GGH} - 2 I_{\hat{F}} - \frac{1}{6}I_{FF} + I_{F R} 
- \frac{1}{18} I_{F\nabla F} - \frac{5}{18}I_{F D F} - I_{\hat{h}\hat{F}} \,,
}{}
it is straightforward to show that all divergent terms cancel. The final result is
\eq{
\de^{(2)}I_{\rm tot}|_{\rm EOM} = \frac{1}{2\kappa^2} \int\limits_{\partial M} \extd^3 x \,\sqrt{-\gamma}
   \left[ \frac{3}{2} \bup{0}\bij{3} + \frac{9}{4} \big(\bup{0}\gij{3} - \hup\bij{3} \big) \right]\, .
}{}

\section{Lightlike Modes}\label{app:llm}

To obtain the lightlike modes we solve \eqref{eq:massive} with the ansatz $\psi_{\mu\nu}(m)=e^{iE(t+x)} \tilde{\psi}_{\mu\nu}(p_i;y)$
and the gauge condition $\nabla^\mu\psi_{\mu\nu}(m)=0$. Going to Gaussian normal coordinates and choosing
combinations of log and Einstein modes such that all $\hij$ are zero we find the following log-modes:
\begin{align}
 \psi^{\rm log1}_{\mu\nu}=\,&e^{i E (t+x^1)}\,\frac{c3\,\log y+c4\, y^3 \log y}{y^2}\,
               \left(\begin{matrix} 0 & 0 & 1 \\ 0 & 0 & 1 \\ 1 & 1 & 0 \end{matrix}\right) \,,\\
 \psi^{\rm log2}_{\mu\nu}=\,&e^{i E (t+x^1)}\,\frac{c9\,\log y+c10\, y^3 \log y}{y^2}\,
               \left(\begin{matrix} 1 & 1 & 0 \\ 1 & 1 & 0 \\ 0 & 0 & 0 \end{matrix}\right) \,.
\end{align}
These are vector modes orthogonal to our chosen $p_i$. Three of the other log-modes turn out
to be Proca modes, their respective Einstein modes are pure gauge modes. Note that there are no explicit
logarithms, but they are identified as log-modes by the equations of motion:
\begin{align}
 \psi^{\rm Proca1}_{\mu\nu}=\,&e^{i E (t+x^1)}\,c2\, y\,\big(1+\frac{y^2 E^2}{25}\big)
               \left(\begin{matrix} 0 & 0 & 1 \\ 0 & 0 & 1 \\ 1 & 1 & 0 \end{matrix}\right) \,,\\
 \psi^{\rm Proca2}_{\mu\nu}=\,&e^{i E (t+x^1)}\,c6\,y\,\Big[
               \left(\begin{matrix} 1 & 0 & 0 \\ 0 & 1 & 0 \\ 0 & 0 & 0 \end{matrix}\right) 
                             +\frac{2 y^2 E^2}{25}
               \left(\begin{matrix} 2 & 0 & 0 \\ 0 & -2 & 0 \\ 0 & 0 & -1 \end{matrix}\right) 
                             +\frac{4 y^4 E^4}{1225}
               \left(\begin{matrix} 1 & 1 & 0 \\ 1 & 1 & 0 \\ 0 & 0 & 0 \end{matrix}\right) \Big] \,,\\
 \psi^{\rm Proca3}_{\mu\nu}=\,&e^{i E (t+x^1)}\,c8\,y\,\Big[
               \left(\begin{matrix} -1 & 0 & 0 \\ 0 & 1 & 0 \\ 0 & 0 & -2 \end{matrix}\right)
                             -\frac{2y^2 E^2}{25}
               \left(\begin{matrix} 1 & 1 & 0 \\ 1 & 1 & 0 \\ 0 & 0 & 0 \end{matrix}\right) \Big]\,.
\end{align}
Finally, we also have
\begin{align}
 \psi^{\rm log3}_{\mu\nu}=\,&e^{i E (t+x^1)}\frac{c1}{y^2}\Big[\log y
               \left(\begin{matrix} 0 & 0 & 1 \\ 0 & 0 & 0 \\ 1 & 0 & 0 \end{matrix}\right)
                             -\frac{y^2 E^2}{4}(3+2\log y)
               \left(\begin{matrix} 0 & 0 & 1 \\ 0 & 0 & 1 \\ 1 & 1 & 0 \end{matrix}\right) \Big] \,,\\
 \psi^{\rm log4}_{\mu\nu}=\,&e^{i E (t+x^1)}\frac{c7}{y^2}\Big[\log y
               \left(\begin{matrix} -1 & 0 & 0 \\ 0 & 1 & 0 \\ 0 & 0 & -2 \end{matrix}\right)
                             -\frac{y^2 E^2}{2}(3+2\log y)
               \left(\begin{matrix} 1 & 1 & 0 \\ 1 & 1 & 0 \\ 0 & 0 & 0 \end{matrix}\right) \Big]\,,
\end{align}
and
\begin{align}
 \psi^{\rm log5}_{\mu\nu}=\,&e^{i E (t+x^1)}\frac{c5}{y^2}\Big[\log y
               \left(\begin{matrix} 1 & 0 & 0 \\ 0 & 1 & 0 \\ 0 & 0 & 0 \end{matrix}\right)
                             +\frac{y^2 E^2}{4}(2+\log y)
               \left(\begin{matrix} -1 & 0 & 0 \\ 0 & 1 & 0 \\ 0 & 0 & -2 \end{matrix}\right) \\
                     &\qquad\qquad        +\frac{y^2 E^2}{4}\log y
               \left(\begin{matrix} -1 & 0 & 0 \\ 0 & 1 & 0 \\ 0 & 0 & 0 \end{matrix}\right)
                             +\frac{y^4 E^4}{32}(5-4\log y)
               \left(\begin{matrix} 1 & 1 & 0 \\ 1 & 1 & 0 \\ 0 & 0 & 0 \end{matrix}\right) \Big]\,.
\end{align}
The corresponding 'simple' Einstein modes read
\begin{align}
 \psi^{\rm Einst1}_{\mu\nu}=\,&e^{i E (t+x^1)}\frac{c3+y^3\,c4}{y^2}\,
               \left(\begin{matrix} 0 & 0 & 1 \\ 0 &0 &1 \\ 1 &1 &0 \end{matrix}\right) \,,\\
 \psi^{\rm Einst2}_{\mu\nu}=\,&e^{i E (t+x^1)}\frac{c9+y^3\,c10}{y^2}\,
               \left(\begin{matrix} 1 & 1 & 0 \\ 1 &1 &0 \\ 0 &0 &0 \end{matrix}\right) \,.
\end{align}
The Einstein modes corresponding to the log-modes $\psi^{\rm log3}$, $\psi^{\rm log4}$ and $\psi^{\rm log5}$
are
\begin{align}
 \psi^{\rm Einst3}_{\mu\nu}=&\, e^{i E (t+x^1)}\frac{c1}{y^2}\Big[
               \left(\begin{matrix} 0 & 0 & 1 \\ 0 &0 &0 \\ 1 &0 &0 \end{matrix}\right)
                             -\frac{y^2 E^2}{2}
               \left(\begin{matrix} 0 & 0 & 1 \\ 0 &0 &1 \\ 1 &1 &0 \end{matrix}\right) \Big]\,,\\
 \psi^{\rm Einst4}_{\mu\nu}=&\, e^{i E (t+x^1)}\frac{c7}{y^2}\Big[
               \left(\begin{matrix} -1 & 0 & 0 \\ 0 &1 &0 \\ 0 &0 &-2 \end{matrix}\right)
                             +y^2 E^2
               \left(\begin{matrix} 1 & 1 & 0 \\ 1 &1 &0 \\ 0 &0 &0 \end{matrix}\right) \Big]\,,\\
 \psi^{\rm Einst5}_{\mu\nu}=&\, e^{i E (t+x^1)}\frac{c5}{y^2}\Big[
               \left(\begin{matrix} 1 & 0 & 0 \\ 0 &1 &0 \\ 0 &0 &0 \end{matrix}\right)
                             +\frac{y^2 E^2}{2}
               \left(\begin{matrix} -1 & 0 & 0 \\ 0 &1 &0 \\ 0 &0 &-1 \end{matrix}\right) 
                             -\frac{y^4 E^4}{8}
               \left(\begin{matrix} 1 & 1 & 0 \\ 1 &1 &0 \\ 0 &0 &0 \end{matrix}\right) \Big]\,.
\end{align}
The gauge modes should be constructed from
\begin{align}
 \psi^{\rm Egauge1}_{\mu\nu}=\,& e^{i E (t+x^1)} \frac{E}{y^2} \,
               \left(\begin{matrix} 2 & 1 & 0 \\ 1 &0 &0 \\ 0 &0 &0 \end{matrix}\right)\,, \qquad
 \psi^{\rm Egauge2}_{\mu\nu}= e^{i E (t+x^1)} \frac{E}{y^2} \,
               \left(\begin{matrix} 0 & 1 & 0 \\ 1 &2 &0 \\ 0 &0 &0 \end{matrix}\right)\,, \\
 \psi^{\rm Egauge3}_{\mu\nu}=\,& e^{i E (t+x^1)} \frac{E}{y^2} \,
               \left(\begin{matrix} 0 & 0 & 1 \\ 0 &0 &1 \\ 1 &1 &0 \end{matrix}\right)\,, \\
 \psi^{\rm Egauge4}_{\mu\nu}=\,& e^{i E (t+x^1)} \frac{1}{y^2} \Big[
               \left(\begin{matrix} 1 & 0 & 0 \\ 0 &-1 &0 \\ 0 &0 &-1 \end{matrix}\right)
                              +\frac{y^2 E^2}{2}
               \left(\begin{matrix} 1 & 1 & 0 \\ 1 &1 &0 \\ 0 &0 &0 \end{matrix}\right)\Big] \,.
\end{align}
As mentioned in the main text all lightlike modes are power series in $y$ (some multiplied by logarithms). Requiring
non-singularity of the modes at $y\to\infty$ kills all the $g^{(3)}_{ij}$ and $b^{(3)}_{ij}$ parts. Therefore,
the lightlike modes do not contribute to the correlators. 

\providecommand{\href}[2]{#2}\begingroup\raggedright\endgroup

% \bibliographystyle{fullsort}
% \bibliography{4dloggrav}

\end{document}